\documentclass[prb,aps,twocolumn,eqsecnum,showpacs,superscriptaddress,amsmath,amssymb]{revtex4}

\usepackage{graphicx}
\usepackage{psfrag}
\usepackage{epsfig}
\usepackage{color}
\usepackage{subfigure}
\definecolor{dred}{rgb}{0.7,0.0,0.0}
\usepackage{dcolumn}
\usepackage{bm}

\begin{document}

\title { Orbitally induced string formation in the spin-orbital polarons
}

\author {     Krzysztof Wohlfeld }
\affiliation{ Max-Planck-Institut f\"ur Festk\"orperforschung,
              Heisenbergstrasse 1, D-70569 Stuttgart, Germany }
\affiliation{ Marian Smoluchowski Institute of Physics, Jagellonian
              University, Reymonta 4, PL-30059 Krak\'ow, Poland }

\author {     Andrzej M. Ole\'{s} }
\affiliation{ Max-Planck-Institut f\"ur Festk\"orperforschung,
              Heisenbergstrasse 1, D-70569 Stuttgart, Germany }
\affiliation{ Marian Smoluchowski Institute of Physics,
              Jagellonian University, Reymonta 4,
              PL-30059 Krak\'ow, Poland }

\author {     Peter Horsch }
\affiliation{ Max-Planck-Institut f\"ur Festk\"orperforschung,
              Heisenbergstrasse 1, D-70569 Stuttgart, Germany }

\date{\today}

\begin{abstract}
We study the spectral function of a single hole doped into the $ab$
plane of the Mott insulator LaVO$_3$, with antiferromagnetic (AF)
spin order of $S=1$ spins accompanied by alternating orbital (AO)
order of active $\{d_{yz},d_{zx}\}$ orbitals. Starting from the respective
$t$-$J$ model, with spin-orbital superexchange and effective
three-site hopping terms, we derive the polaron Hamiltonian and
show that a hole couples simultaneously to the collective
excitations of the AF/AO phase, magnons and orbitons. Next, we
solve this polaron problem using the self-consistent Born
approximation and find a stable quasiparticle solution --- a
spin-orbital polaron. We show that the spin-orbital polaron
resembles the orbital polaron found in $e_g$ systems, as e.g. in
K$_2$CuF$_4$ or (to some extent) in LaMnO$_3$, and that the hole
may be seen as confined in a string-like potential. However, the
spins also play a crucial role in the formation of this polaron
--- we explain how the orbital degrees of freedom: (i) confine the
spin dynamics acting on the hole as the classical Ising spins, and
(ii) generate the string potential which is of the joint
spin-orbital character. Finally, we discuss the impact of the
results presented here on the understanding of the phase diagrams
of the lightly doped cubic vanadates.
\end{abstract}

\pacs{71.10.Fd, 72.10.Di, 72.80.Ga, 79.60.-i}

\maketitle

\section{INTRODUCTION}
\label{sec:introduction}

A single hole doped into the half-filled Mott insulating ground
state with antiferromagnetic (AF) order does not move freely as
its motion disturbs the spin order.\cite{Bul68} Instead, it
couples to the collective (delocalized) excitations of the AF
ordered phase (magnons), and it propagates through the lattice
together with a "cloud" of magnons. \cite{Kan89} Thereby the
energy scale of the "coherent" hole propagation is strongly
renormalized and is given by the AF superexchange constant $J$. In
this way a quasiparticle is formed which is frequently called in
the literature a {\it spin polaron}.\cite{Mar91} Actually, such a
phenomenon is not only of theoretical importance but also
describes a realistic situation found in the photoemission spectra
of the parent compounds of the high-$T_c$ cuprates, as e.g. in
Sr$_2$CuO$_2$Cl$_2$.\cite{Dam03}

\begin{figure}[b!]
\includegraphics[width=0.3\textwidth]{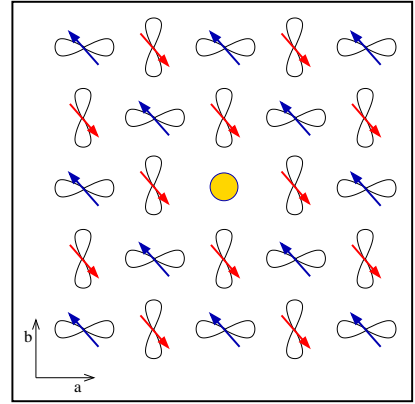}
\caption{(Color online) Artist's view of a single hole introduced
into the spin and orbitally ordered $ab$ plane of LaVO$_3$ which
leads to the spin-orbital polaron considered in the paper. The
electrons occupy the $d_{yz}$ and $d_{zx}$ degenerate orbitals
forming the classical AO state (their projections along the $a$
and $b$ axis are shown) whereas the electron spins alternate on
the neighboring sites forming the classical AF state. 
Only the spin $1/2$ of the electrons in the $d_{yz}$ and $d_{zx}$ orbitals
are shown. At each site an additional electron occupies the $d_{xy}$
orbital (not shown). Due to Hund's coupling
the two electron spins form a total spin
$S=1$, while at the site of the hole $S=1/2$.}
\label{fig:1}
\end{figure}

A more complex situation can occur in the systems with partly
filled degenerate orbitals, where a doped hole may not only couple
to magnons but may also couple to the crystal-field excitations.
\cite{Zaa93} Then we may for example arrive at a problem shown
schematically in Fig. \ref{fig:1}, which is studied in this paper
and discussed in more detail below, or end up in two somewhat
simpler, though still challenging, situations where at least the
coupling to the magnons could be neglected.
To get a better
insight into the {\it orbital physics\/} we give a brief overview
of these two latter cases in the next two paragraphs.

First, we recall an example of the $ab$ planes of LaMnO$_3$ which
have ferromagnetic (FM) spin order and alternating orbital (AO)
order of $e_g$ orbitals in the ground state.\cite{Fei99} It has
been shown\cite{Bri00} that although the hole introduced into such
a state does not disturb the FM spin order, it couples to the
collective excitations of the AO state (orbitons). Here again a
quasiparticle is formed which is called this time an {\it orbital
polaron}.\cite{Kil99} It should be noted, however, that the
orbital polaron has an even smaller bandwidth than the spin
polaron, \cite{Bri00} as the orbitons are in general much less
mobile than the magnons (or even immobile) due to the lack of the
SU(2) symmetry in the orbital systems,\cite{Fei05} and almost
directional Ising-like superexchange. \cite{Fei99,vdB99} Actually,
one can understand the hole motion in this case in terms of the
string picture:\cite{Wro08} The hardly mobile orbitons cannot
repair the string of the misaligned orbitals in the AO state,
which arises by the hole motion. Thus, it is the hole which has to
return to the original site and cure the defects by retracing its
path, unless it propagated due to small off-diagonal hopping in an
$e_g$ system \cite{Bri00} and no defects were created on its path
(the latter process also contributes to the above mentioned very
small bandwidth of the orbital polaron).

Second, only recently an even more extreme situation of the system
with orbital order was investigated:\cite{Dag08,Woh08} The case of
a hole doped into the plane with FM spin order accompanied by the
$t_{2g}$ AO order (which could correspond not only to the hole
introduced into the ordered ground state of Sr$_2$VO$_4$ with
$t_{2g}$ orbitals but also, surprisingly, to those of K$_2$CuF$_4$
or Cs$_2$Ag$_4$ with $e_g$ active orbitals, see Ref.
\onlinecite{Woh08}). Also here a quasiparticle is formed due to
the dressing of a hole by the collective excitations of the ground
state with AO order. However, due to the specific $t_{2g}$ orbital
symmetries the orbitons are not mobile at all, the off-diagonal
hopping is prohibited, and the quasiparticle acquires a finite
bandwidth only due to the frequently neglected three-site
terms.\cite{Dag08} Thus, the string picture dominates the
character of the $t_{2g}$ orbital polarons even more than in the
case of systems with $e_g$ orbital degrees of freedom.

Finally, after this brief overview, we turn to the problem
investigated in the present paper: the motion of a single hole
introduced into the ground state with coexisting AF order of $S=1$
spins and $t_{2g}$ AO order, shown schematically in Fig.
\ref{fig:1}. This situation corresponds to a single hole doped
into the $ab$ plane of LaVO$_3$.\cite{Kha01} It differs from all
of the three (one --- purely spin and the other two --- purely
orbital) cases described above due to the coexisting
AF/AO order. Neither spin nor the orbital background is then
transparent for the propagating hole, and the hole has to couple
{\it both\/} to the magnons and orbitons simultaneously (see Sec.
\ref{sec:coupling}). To our knowledge, the mechanism and physical
consequences of this coupling have never been studied
before.\cite{note0}

Furthermore, the coexistence of the AO and AF order is an
extremely rare case in nature as it formally violates\cite{Ole06}
the Goodenough-Kanamori rules that state complementary spin and
orbital order in the ground state, i.e., either FM spin coexisting
with AO order, or AF spin coexisting with FO order --- thus, it is
interesting to investigate how a hole can propagate in a state
which does not fulfil this rule and by its nature contains more
quantum fluctuations. Therefore, the main goals of this paper are
to verify: (i) {\it under which conditions}, if at all, {\it a
quasiparticle can be formed} in such an AF phase with AO order,
(ii) if a quasiparticle is formed,{\it what are its properties}
and whether the string picture (which is typical for orbital
physics) can explain them, and finally (iii) {\it what is the role
of spins and orbitals in the possible formation of the string} in
the spin-orbital polarons. By working out the answers to the above
questions we would like to predict the main features of the
photoemission spectra of the cleaved samples of LaVO$_3$ with
polarization corresponding to the $ab$ planes.

Before we move on to present the answers to these questions,
let us note another motivation for studying the above problem.
It concerns the phase diagram of the lightly doped cubic vanadate
La$_{1-x}$Sr$_x$VO$_3$.\cite{Miy00} It appears that in this
strongly correlated compound the $C$-AF (AF $ab$ planes with FM
order along the $c$ direction) coexisting with $G$-AO (AO order in
all three cubic directions) order Mott insulating phase is not
only stable at $x=0$ but also persists to a relatively high value
of doping $x=0.178$.\cite{Miy00} Furthermore, the $C$-AF phase
remains stable up to an even higher value of $x=0.26$, although in
this regime the insulating and orbital ordered phase has already
disappeared.\cite{Miy00} As in the ionic picture $x$ stands for
the fraction of holes doped into the $d$ orbitals of vanadium ions
(where a nominal valence upon doping changes as $d^{2-x}$), it
remains a challenge to explain why the ordered and insulating
states persist to such high doping level.\cite{Woh06} Besides,
somewhat similar phase diagrams have been observed in other doped
cubic vanadates,\cite{Fuj05} such as Pr$_{1-x}$Ca$_x$VO$_3$,
Nd$_{1-x}$Sr$_x$VO$_3$ or even to some extent in
Y$_{1-x}$Ca$_x$VO$_3$ although in all these cases the lattice
(Jahn-Teller and GdFeO$_3$-like) distortions contribute
significantly to their physical properties, similar to the undoped
parent $R$VO$_3$ (with $R$=Pr,Nd,Y) compounds.\cite{Hor08} Here to
avoid further complications due to additional orbital interactions
imposed by the lattice, we would like to concentrate ourselves on
the spin-orbital polarons in (almost) undistorted compound
LaVO$_3$.

Actually, one should in principle be able to understand some of
the features observed in the phase diagram of the doped cubic
vanadates by comparing them with those of the high-$T_c$ cuprates
or the colossal magnetoresistive manganites. However, such a
comparison only further emphasizes the lack of theoretical
understanding of the doped cubic vanadates. First, in the
high-$T_c$ cuprates, such as La$_{2-x}$Sr$_x$CuO$_4$, the AF order
disappears very quickly with doping $x$, i.e., already for
$x\sim0.02$.\cite{Ima98} This is despite the fact that the value
of the superexchange constant $J$ is relatively high there which
naturally leads to larger magnetic energy in the cuprates than in
the vanadates. This suggests that, among other factors, it is the
orbital dynamics which could be responsible for the totally
different behavior of these two classes of compounds upon hole
doping. Second, a similar conjecture can be drawn from the
comparison between the vanadates and the manganites. In the latter
ones, e.g. in La$_{1-x}$Sr$_x$MnO$_3$, the AO orbital Mott
insulating state is stable up to $x\sim 0.18$,\cite{Bri04} i.e.,
almost to the same doping level as in La$_{1-x}$Sr$_x$VO$_3$.
However, La$_{1-x}$Sr$_x$MnO$_3$ has FM planes but
La$_{1-x}$Sr$_x$VO$_3$ has AF planes -- this again indicates that
it is the orbital dynamics which governs the behavior of holes in
the doped cubic vanadates.

Hence, we arrive at the following paradox. On the one hand, it is
obvious from the theory that any meaningful study of the doped
perovskite vanadates should take into account the spin and orbital
degrees of freedom on equal footing. On the other hand, we see
from the above conjecture from the experiment that these are the
orbital degrees which are, to large extent, responsible for the
stability of the observed phases of the lightly doped cubic
vanadates. Thus, another aim of the paper is to shed some light on
this apparent paradox. Certainly, the studies presented below
concern only the case of just a single hole doped into the
half-filled compound and, furthermore, they investigate only the
situation in the two-dimensional (2D) model focusing on the $ab$
planes. Nevertheless, we hope to understand the generic role of
the spin and orbital degrees of freedom in the hole doped cubic
vanadates. Besides, we note that complementary studies presented
in Ref. \onlinecite{Ish05} revealed the role of the AO and FM
order stable along the third (not studied here) direction in the
doped La$_{1-x}$Sr$_x$VO$_3$, and explained the differences
between the doped Y$_{1-x}$Ca$_x$VO$_3$ and La$_{1-x}$Sr$_x$VO$_3$
but, by the very nature of that one-dimensional model, could not
address the problems mentioned above. We also stress that,
contrary to the problem solved in this paper, the hole moving
along this third $c$ cubic direction couples {\it only} to {\it
one} type of excitations: either orbitons in the lightly doped
$C$-AF phase of La$_{1-x}$Sr$_x$VO$_3$, or magnons in the very
lightly doped $G$-AF phase of Y$_{1-x}$Ca$_x$VO$_3$. \cite{Ish05}

After a thorough discussion of the motivation standing behind the
studies presented in the paper let us now concentrate on the model
and the theoretical method used in this work. We use the 2D
$t$-$J$ model for the V$^{3+}$ ions in the $t_{2g}^2$
configuration with degenerate $t_{2g}$ orbitals, and the
superexchange given in Refs. \onlinecite{Kha01,Ole05}, as {\it the
model} which describes the physical properties of a hole in the
$ab$ plane of the lightly doped cubic vanadates. Since any $t$-$J$
model which contains Ising-type interactions should be
supplemented by the three-site effective hopping terms in order to
provide a possibility of coherent hole
propagation,\cite{Dag08,Woh08} we derive these terms and add them
to the $t$-$J$ model (such a model is then called the
strong-coupling model). Next, we solve the strong-coupling model
in the case of a single hole doped into the ground state of a Mott
insulator by reducing it to the polaron-like model and introducing
the self-consistent Born approximation (SCBA).\cite{Mar91}
Although the final solutions (spectral functions) are obtained
numerically on a finite mesh of the momentum ${\bf k}$ points, the
SCBA method is known to be very accurate to study such problems,
\cite{Dag94, Mar91} and is largely independent of the size of the
clusters\cite{Mar91} on which the calculations are performed.

The paper is organized as follows. We start with introducing the
spin-orbital $t$-$J$ model in Sec. \ref{sec:spin-orbital} which is
supplemented by the crucial three-site terms for the spin-orbital
model, see Appendices \ref{app:free} and \ref{app:2magnons} for
detailed discussion and derivation. We reduce the derived
strong-coupling model ($t$-$J$ model with three-site terms) to the
spin-orbital polaron problem in Secs. \ref{sec:undoped}
-\ref{sec:free} and discuss its parameters in Sec.
\ref{sec:parameters}. Next, we formulate the solution of the model
using the analytic approach within the SCBA method, see Sec.
\ref{sec:results}. In Sec. \ref{sec:num} we present the numerical
results and discuss the properties of the quasiparticle states.
Finally, in Sec. \ref{sec:discussion} we analyze the composite
interplay of spin and orbital degrees of freedom in the formation
of the spin-orbital polaron whereas in Sec. \ref{sec:conclusions}
we present our conclusions. We adopt a convention throughout the
paper [despite the introductory part (Sec. \ref{sec:introduction})
and Appendix \ref{app:free}], that the red (blue) color in the
figures denotes features associated solely with spins (orbitals).

\section{THE MODEL}
\label{sec:model}

\subsection{The spin-orbital $t$-$J$ model with three-site terms}
\label{sec:spin-orbital}

The Hamiltonian of the spin-orbital $t$-$J$ model with the
three-site terms (called also the strong-coupling
spin-orbital Hamiltonian), relevant for the planes of lightly
doped cubic vanadates, consists of three terms,
\begin{equation}\label{eq:str-coupl}
H=H_t+H_J+H_{3s},
\end{equation}
where the last part $H_{3s}$ stands for the three-site terms.
The first term in Eq. (\ref{eq:str-coupl}), which describes the
hopping of the electrons in the constrained Hilbert space, i.e.,
in the space with singly occupied or doubly occupied sites,
follows in a straightforward way from the unconstrained hopping of
electrons residing in the $t_{2g}$ orbitals. The latter one has to
fulfill the $t_{2g}$ orbital symmetries meaning that along each
bond in the cubic direction only two out of three $t_{2g}$
orbitals are active. \cite{Kha01,Har04} This means that in the
$ab$ plane, which is under consideration here, electrons in
$d_{yz}\equiv a$ ($d_{zx}\equiv b$) orbitals can hop only along
the $b$ ($a$) direction. Besides, the $d_{xy}$ orbital does not
contribute to hopping elements as it lies lower in energy and is
always occupied by one electron in the half-filled and lightly
hole-doped regime. \cite{Fuj06} Hence, we arrive at the kinetic
$\propto t$ part of Hamiltonian (\ref{eq:str-coupl}),
\begin{equation}\label{eq:ht}
H_t = -t \sum_{{\bf i},\sigma} {\cal P}\left( \tilde{b}^\dag_{{\bf i}\sigma}
\tilde{b}_{{\bf i}+\hat{\bf a},\sigma}
+\tilde{a}^\dag_{{\bf i}\sigma}\tilde{a}_{{\bf i}+\hat{\bf b},\sigma} + {\rm
H.c.}\right) {\cal P}.
\end{equation}
Here the constrained operators
\begin{eqnarray}
\tilde{b}^\dag_{{\bf i}\sigma}&=&b^\dag_{{\bf i}\sigma}(1-n_{{\bf i}b\bar{\sigma}})
(1-n_{{\bf i}a\bar{\sigma}})(1-n_{{\bf i}a\sigma}), \\
\tilde{a}^\dag_{{\bf i}\sigma}&=&a^\dag_{{\bf i}\sigma}(1-n_{{\bf i}a\bar{\sigma}})
(1-n_{{\bf i}b\bar{\sigma}})(1-n_{{\bf i}b\sigma}),
\end{eqnarray}
mean that the hopping is allowed only in the restricted Hilbert
space with not more than one $\{a,b\}$ electron at each site ${\bf i}$
($\bar{\sigma}$ stands for the spin component opposite to
$\sigma$). Besides, since the Hund's coupling is large ($J_H\gg
t$) in the cubic vanadates,\cite{Miz96} we project the final
states resulting from the electron hopping onto the high spin
states, which is denoted by the ${\cal P}$ operators in Eq.
(\ref{eq:ht}).

The middle term in Eq. (\ref{eq:str-coupl}) describes the
spin-orbital superexchange in cubic vanadates\cite{Kha01} --- here
we use the notation of Ref. \onlinecite{Ole05}, see Eqs.
(6.5)-(6.7). It reads
\begin{align}\label{eq:hj}
H_J=H^{(1)}_J+H^{(2)}_J+H^{(3)}_J,
\end{align}
where the contribution due to different excitations in the
multiplet structure of vanadium ions are:\cite{Kha04}
\begin{align}\label{eq:hj123}
H^{(1)}_J&=-\frac{1}{6}Jr_1 \sum_{\langle {\bf i}{\bf j}\rangle}
\left({\bf S}_{\bf i} \cdot {\bf S}_{\bf j} + 2\right)
\left(\frac{1}{4}-T^z_{\bf i} T^z_{\bf j}\right), \\
H^{(2)}_J&=\frac{1}{16}J \sum_{\langle {\bf i}{\bf j} \rangle}
\left({\bf S}_{\bf i} \cdot {\bf S}_{\bf j} - 1\right) \left(\frac{19}{6}
\mp T^z_{\bf i} \mp T^z_{\bf j} -\frac{2}{3} T^z_{\bf i} T^z_{\bf j}\right), \\
H^{(3)}_J&=\frac{1}{16}J r_3 \sum_{\langle {\bf i}{\bf j} \rangle}
\left({\bf S}_{\bf i} \cdot {\bf S}_{\bf j}-1\right) \left(\frac{5}{2} \mp
T^z_{\bf i} \mp T^z_{\bf j} + 2T^z_{\bf i} T^z_{\bf j}\right).
\end{align}
Here: ${\bf S}_{\bf i}$ is a spin $S=1$ operator,
$T^z_{\bf i}=(\tilde{n}_{{\bf i}b}-\tilde{n}_{{\bf i}a})/2$ is a pseudospin $T=1/2$
operator, and $J=4t^2/U$ is the superexchange constant, with $U$
being the intraorbital Coulomb element on a vanadium V$^{2+}$ ion.
As in the spin $t$-$J$ model,\cite{Cha77} we assume that the
hopping $t$ is small, i.e., $t\ll U$ whose assumption is well
satisfied by the realistic parameters in the cubic
perovskites.\cite{Kha01} The factors $r_1=1/(1-3\eta)$ and
$r_3=1/(1+2\eta)$, where
\begin{equation}
\label{eta} \eta=\frac{J_H}{U}\,,
\end{equation}
account for the Hund's coupling $J_H$ and originate from the
multiplet structure of the $d^3$ excited states characterized by
the various allowed spin and orbital configurations.\cite{Kha01}
Let us note, that superexchange Hamiltonian (\ref{eq:hj}) was
derived \cite{Ole05} with the assumption that the $d_{xy}$ orbital
was singly occupied at each Vanadium ion (see discussion above).
Besides, in principle Hamiltonian (\ref{eq:hj}) was originally
derived for the undoped case and should be modified for the doped
case by adding the superexchange interactions due to the existence
of the $d^1_{\bf i} d^1_{\bf j} $ and $d^1_{\bf i}d^2_{\bf j}$
nearest neighbor configurations. However, in the discussed here
small doping regime, and particularly for a single hole, the
modification of magnon and orbiton excitations due to these terms
should be very small and we will neglect them.

The last term in Eq. (\ref{eq:str-coupl}) denotes the three-site
terms which, to our knowledge, have not been derived before for
the case of the $d^2$ systems with spin and orbital degrees of
freedom. These terms, although frequently neglected, are crucial
for a faithful representation of the Hubbard model in the
strong-coupling regime,\cite{Esk94} and can play an important role
in the coherent hole propagation in orbital systems.\cite{Dag08, Woh08}
However, in the present case the derivation of all possible
three-site terms is relatively tedious and leads to a complex
expression. Fortunately, it was shown in Ref. \onlinecite{Woh08}
(for orbital systems) and in Ref. \onlinecite{Bal95} (for spin
systems) that the only three-site terms which occurred to be
relevant for the lightly doped systems were these which did not
contribute to the coupling between hole and orbital or spin
excitation but merely contributed to the free hole motion (see
Sec. \ref{sec:free}). Nevertheless, we discuss the possible role
of the neglected three-site terms in Appendix \ref{app:2magnons}.
Then, the derived in Appendix \ref{app:free} three-site terms
which could possibly contribute to the free hole motion read:
\begin{eqnarray}\label{eq:h3s}
H_{3s}\!\! &=&\!\! - \frac{1}{12} J \left(r_1+2\right)
\sum_{{\bf i},\sigma} {\cal P} \left( \tilde{b}^\dag_{{\bf i}-\hat{\bf a},\sigma}
\tilde{n}_{{\bf i}a\bar{\sigma}}
\tilde{b}_{{\bf i}+\hat{\bf a},\sigma}+ {\rm H. c.} \right) {\cal P}  \nonumber \\
&-&  \frac{1}{12} J \left(r_1+2\right) \sum_{{\bf i},\sigma} {\cal P}
\left( \tilde{a}^\dag_{{\bf i}-\hat{\bf b},\sigma}
\tilde{n}_{{\bf i}b\bar{\sigma}}
\tilde{a}_{{\bf i}+\hat{\bf b},\sigma} + {\rm H. c.} \right) {\cal P}.\nonumber \\
\end{eqnarray}
Note that these terms are $\propto J$ and hence are of the same
order in $t^2/U$ as superexchange terms (\ref{eq:hj}).

In the next four sections we reduce the model Eq.
(\ref{eq:str-coupl}) to the polaron problem following Ref.
\onlinecite{Mar91}. This would enable us to solve the model Eq.
(\ref{eq:str-coupl}) using the SCBA method.

\subsection{Undoped case: low energy excitations}
\label{sec:undoped}

It was shown in Ref. \onlinecite{Kha01} that the ground
state of the 3D spin-orbital superexchange model (\ref{eq:hj123})
is characterized by the classical phase with $C$-AF spin
order and $G$-AO orbital order. Thus, we start from the above
{\it undoped} ground state of the Hamiltonian Eq. (\ref{eq:str-coupl})
with classical (N\'eel) order. As usual, this is not the
eigenstate of the Hamiltonian and thus the full description of the
system has to take into account the quantum fluctuations around
such a classical ground state. Below, we will calculate them,
together with the low energy (magnon and orbiton) excited states,
employing the linear spin wave (LSW) and linear orbital wave (LOW)
approximation, following Ref. \onlinecite{Ole07}.

First, in the classical state we introduce two sublattices $A$ and
$B$ such that: (i) all $a$ ($b$) orbitals are occupied in the
perfect AO state in sublattice $A$ ($B$), and (ii) spins pointing
"downwards" ("upwards") are located on sublattice $A$ ($B$), as
shown schematically in Fig. \ref{fig:1}. Next we rotate spins and
orbitals (pseudospins) on sublattice $A$ so that all the spins and
pseudospins are aligned in the considered $ab$ plane and take
positive values now.

Second, we introduce Schwinger bosons $t$ and $f$ for orbitals and
spins such that:
\begin{align}
T^z_{\bf i}&=\frac{1}{2} (t^{\dag}_{{\bf i}b} t_{{\bf i}b}-t^{\dag}_{{\bf i}a} t_{{\bf i}a})\,, \\
S^z_{\bf i} & =\frac{1}{2} (f^\dag_{{\bf i}\uparrow} f_{{\bf i}\uparrow}-
f^\dag_{{\bf i}\downarrow} f_{{\bf i}\downarrow})\,, \\
S^{+}_{\bf i} & =f^\dag_{{\bf i}\uparrow}f^{}_{{\bf i}\downarrow}\,, \\
S^{-}_{\bf i} & = f^\dag_{{\bf i}\downarrow}f^{}_{{\bf i}\uparrow}\,,
\end{align}
with the constraints
\begin{equation}
\sum_{\gamma=a,b} t^\dag_{{\bf i}\gamma}t_{{\bf i}\gamma} = 1\,, \hskip .7cm
\sum_{\sigma=\uparrow,\downarrow}
f^\dag_{{\bf i}\sigma}f^{}_{{\bf i}\sigma}=2\,.
\end{equation}
Third, we transform the Schwinger boson operators into
the Holstein-Primakoff bosons $\alpha$ and $\beta$:
\begin{align}
t^\dag_{{\bf i}b}&=\sqrt{1-t^\dag_{{\bf i}a} t^{}_{{\bf i}a}}
\equiv\sqrt{1-\beta^\dag_{\bf i}\beta^{}_{\bf i}}, \\
t^\dag_{{\bf i}a} &= \beta^\dag_{\bf i}, \\
f^\dag_{{\bf i}\uparrow}&=\sqrt{2-f^\dag_{{\bf i}\downarrow} f^{}_{{\bf i} \downarrow}}
\equiv \sqrt{2-\alpha^\dag_{\bf i} \alpha^{}_{\bf i}}, \\
f^\dag_{{\bf i} \downarrow} &= \alpha^\dag_{\bf i},
\end{align}
where the above constraints are now no longer needed.

Next, we substitute the above transformations into the Hamiltonian
$H_J$ and skip higher order terms (using LSW and LOW
approximation). The latter approximation physically means that the
number of bosons $\alpha_{\bf i}$ and $\beta_{\bf i}$, which describe the
fluctuations around the ordered state, is small. This results in
the effective substitutions:
\begin{align}
T^z_{\bf i}&=\frac{1}{2} - \beta^\dag_{\bf i} \beta_{\bf i}\,, \\
S^z_{\bf i} & =1- \alpha^\dag_{\bf i} \alpha_{\bf i}\,,\\
S^{+}_{\bf i} & = \sqrt{2} \alpha_{\bf i}\,, \\
S^{-}_{\bf i} &= \sqrt{2} \alpha_{\bf i}^\dag\,.
\end{align}

Finally, we introduce Fourier
transformation separately for each sublattice
($N$ is the total number of sites on both sublattices and the lattice constant $a=1$):
\begin{align}
\beta_{{\bf k}A}&=\sqrt{\frac{2}{N}}\sum_{{\bf j} \in A} e^{i{\bf k j}} \beta_{\bf j}\,, \\
\beta_{{\bf k}B}&=\sqrt{\frac{2}{N}}\sum_{{\bf j} \in B} e^{i{\bf k j}} \beta_{\bf j}\,, \\
\alpha_{{\bf k}A}&=\sqrt{\frac{2}{N}}\sum_{{\bf j} \in A} e^{i{\bf k j}} \alpha_{\bf j}\,, \\
\alpha_{{\bf k}B}&=\sqrt{\frac{2}{N}}\sum_{{\bf j} \in B} e^{i{\bf k
j}} \alpha_{\bf j}\,,
\end{align}
we define operators
\begin{align}\label{eq:alpha}
\alpha_{{\bf k}\pm}=(\alpha_{{\bf k}A}\pm \alpha_{{\bf k}B})/\sqrt{2},
\end{align}
and perform the standard Bogoliubov transformation to diagonalize
the magnon part of the Hamiltonian:
\begin{align}
\tilde{\alpha}_{{\bf k}\pm}=u_{\bf k}\alpha_{{\bf k}\pm}
-v_{\bf k}\alpha^\dag_{-{\bf k}\pm},
\end{align}
where
\begin{align}
\label{uv} u_{\bf k}=\sqrt{\frac{1+\nu_{\bf k}}{2\nu_{\bf k}}}\,,
\quad v_{\bf k}=-\mbox{sgn}(\gamma_{\bf k})\sqrt{\frac{1-\nu_{\bf
k}}{2\nu_{\bf k}}}\,,
\end{align}
with $\nu_{\bf k}=\sqrt{1-\gamma^2_{\bf k}}$ and the structure
factor for the square lattice $\gamma_{\bf k}=(\cos k_x+\cos
k_y)/2$.

Then, after neglecting constant terms, the LSW and LOW
Hamiltonian for magnons and orbitons reads:
\begin{align}\label{eq:hjlsw}
{\cal H}_J&=H_{\mbox{\tiny LSW}}+H_{\mbox{\tiny LOW}}, \\
\label{eq:lsw} H_{\mbox{\tiny LSW}}&=J_S\sum_{\bf k } \omega_{\bf
k} (\tilde{\alpha}^\dag_{{\bf k}+} \tilde{\alpha}^{}_{{\bf k}+}
+\tilde{\alpha}^\dag_{{\bf k}-} \tilde{\alpha}^{}_{{\bf k}-} + 1), \\
\label{eq:low} H_{\mbox{\tiny LOW}}&=J_O\sum_{\bf k }
(\beta^\dag_{{\bf k}A} \beta^{}_{{\bf k}A} +\beta^\dag_{{\bf
k}B}\beta^{}_{{\bf k}B}),
\end{align}
where
\begin{equation}\label{eq:jo}
J_O = \eta\,\frac{2-\eta}{(1-3\eta)(1+2\eta)}\,J\,,
\end{equation}
and
\begin{equation}\label{eq:js}
J_S = \frac{1-3 \eta-5 \eta^2}{4 (1-3\eta)(1+2\eta)}\,J\,.
\end{equation}
in agreement with Eq. (6.11) of Ref. \onlinecite{Ole05} and Eq.
(11) of Ref. \onlinecite{Kha04}. Besides the dispersion relation
of the magnons is
\begin{equation}
\omega_{\bf k}=zS\sqrt{1-\gamma^2_{\bf k}}\,,
\end{equation}
where $z=4$ is the coordination number and $S=1$ is the spin
value. Let us note, that in the regime of reasonable values of
$\eta\in [0,0.20]$: $J_O>0$ and $J_S>0$, which means that the
classical ground state has indeed coexisting AF and AO order.
Furthermore, at temperature $T=0$ the considered here
classical 2D AO and AF ground state is stable with respect
to the quantum fluctuations, both in spin and in orbital channel.
In fact, the orbital order is undisturbed by local Ising
excitations, while the quantum AF ground state is modified and the
order parameter is renormalized with magnon
excitations.\cite{Mar91}

\subsection{Doped hole:\\coupling with magnons and orbitons}
\label{sec:coupling}

We expect that a doped hole does not modify significantly the
classical ground state stable for the half-filled case (see above).
This could play a role in the lightly doped regime, but in
the case of one hole in the whole plane such a modification
is negligible and will be neglected below. Instead, the doped
hole may modify its neighborhood by its coupling to the
excitations of the classical ground state --- magnons and
orbitons --- which renormalize the hole motion.
In  order to describe it mathematically, we rewrite $H_t$ (see
below) and $H_{3s}$ (see next section) using similar
transformations as performed in Sec. \ref{sec:undoped}.

First, we rotate spins and pseudospins on sublattice $A$. Next,
we express the electron operators in terms of the
Schwinger bosons introduced in Sec. \ref{sec:undoped}:
\begin{align}
\tilde{a}^\dag_{{\bf i} \sigma} &= \frac{1}{\sqrt{2}}
f^\dag_{{\bf i} \sigma} t^\dag_{{\bf i}a} h^{}_{\bf i}\,, \\
\tilde{b}^\dag_{{\bf i} \sigma} &= \frac{1}{\sqrt{2}} f^\dag_{{\bf i} \sigma}
t^\dag_{{\bf i}b} h^{}_{\bf i}\,,
\end{align}
with the same constraints as in Sec. \ref{sec:undoped}. Note that
the factor $\frac{1}{\sqrt{2}}$ is not added "ad hoc" but
originates from a detailed check of the validity of the above
equations: It should always be present in the case of spin $S=1$
because e.g. when one annihilates two-boson state with the $f$
operator, then a factor $\sqrt{2}$ appears. Due to this factor and
the above constraint on the number of bosons the high spin
projection operators ${\cal P}$ in $H_t$ are no longer needed
(i.e., quantum double exchange \cite{Loo01} factors are implicitly
included in this formalism).

Next, similarly as in Sec. \ref{sec:undoped}, we transform the
Schwinger bosons into the Holstein-Primakoff bosons, skip all
terms containing more than two bosons, perform Fourier
transformation for bosons and (additionally) for holons here,
introduce $\alpha_{{\bf k}\pm}$ operators, and finally perform
Bogoliubov transformation to arrive at the Hamiltonian:
\begin{align}\label{eq:htpolaron}
{\cal H}_t\!=& \frac{zt}{2N}\!\sum_{\bf k, q_1, q_2 } \Big[ M_x
({\bf k, q_1, q_2}) h^\dag_{{\bf k}A}h_{{\bf \bar{k}}B}
(\tilde{\alpha}_{{\bf q}_{1+}} + \tilde{\alpha}_{{\bf q}_{1-}})
\beta_{{\bf q}_2 A}  \nonumber \\
+&M_y ({\bf k, q_1, q_2}) h^\dag_{{\bf k}B}h_{{\bf \bar{k}}A}
(\tilde{\alpha}_{{\bf q}_{1+}} - \tilde{\alpha}_{{\bf q}_{1-}})
\beta_{{\bf q}_2 B}  \nonumber \\
+&M_x ({\bf k, q_1, q_2}) h^\dag_{{\bf k}A}h_{{\bf \bar{k}}B}
(\tilde{\alpha}^\dag_{-{\bf q}_{1+}}-\tilde{\alpha}^\dag_{-{\bf
q}_{1-}})
\beta_{{\bf q}_2 A}  \nonumber \\
+&M_y ({\bf k, q_1, q_2}) h^\dag_{{\bf k}B}h_{{\bf \bar{k}}A}
(\tilde{\alpha}^\dag_{-{\bf q}_{1+}} +\tilde{\alpha}^\dag_{-{\bf
q}_{1-}}) \beta_{{\bf q}_2 B} + \mbox{H.c.} \Big]\,,
\end{align}
where
\begin{equation}
M_\mu ({\bf k, q_1, q_2})= u_{{\bf
q}_1}\gamma_{k_\mu-q_{1\mu}-q_{2\mu}} + v_{{\bf q}_1}
\gamma_{k_\mu-q_{2\mu}}\,,
\end{equation}
with $\mu=x,y$ and $\bar{\bf k}= {\bf k -q_1-q_2}$; the
coefficients $\{u_{{\bf q}_1},v_{{\bf q}_1}\}$ are the standard
Bogoliubov factors (\ref{uv}).

\subsection{Doped hole: the free dispersion}
\label{sec:free}

After performing similar transformations as the ones introduced in
Sec. \ref{sec:coupling} we obtain that the three-site terms, Eq.
(\ref{eq:h3s}), lead to the following Hamiltonian for the holes
\begin{align}\label{eq:h3spolaron}
{\cal H}_{3s} &= \tau  \sum_{\bf k} \left\{\varepsilon_B({\bf k})
h^\dag_{{\bf k}B} h^{}_{{\bf k}B} + \varepsilon_A({\bf k})
h^\dag_{{\bf k}A} h^{}_{{\bf k}A}\right\},
\end{align}
where
\begin{align}\label{eq:tau}
\tau=\frac{1}{4}\frac{1-2\eta}{ 1 - 3 \eta}\ J\,,
\end{align}
and the free dispersion relations for two orbital flavors are:
\begin{eqnarray}
\varepsilon_A({\bf k})&=&2\cos (2 k_y)\,, \\
\varepsilon_B({\bf k})&=&2\cos (2 k_x)\,.
\end{eqnarray}
Note, that, we neglected all of the three-site terms which lead to
the coupling between holes and magnons and/or orbitons. This
approximation is physically well justified since all such terms
would be of the order of $J/4$, i.e., much smaller than the terms
in Eq. (\ref{eq:htpolaron}). See also Appendix \ref{app:2magnons}
for further discussion.

\subsection{Polaron model and its parameters}
\label{sec:parameters}

\begin{figure}
\includegraphics[height=0.25\textwidth]{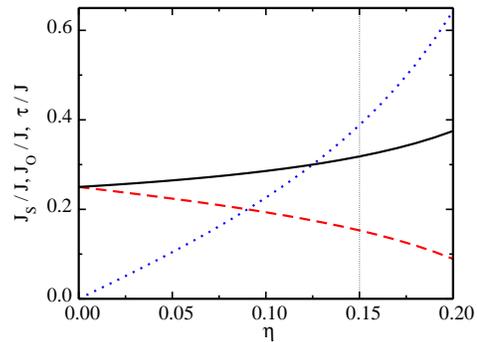}
\caption{(Color online) The three-site hopping term $\tau$
(\ref{eq:tau}) (solid line), the effective spin exchange $J_S$
(\ref{eq:js}) and orbital exchange $J_O$ (\ref{eq:jo}) interaction
(dashed and dotted line) for increasing Hund's exchange coupling
$\eta$ (\ref{eta}). The realistic value of $\eta=0.15$ for
LaVO$_3$ (cf. Ref. \onlinecite{Kha01}) is indicated by the light
vertical dotted line. } \label{fig:2}
\end{figure}

The above considerations demonstrate that in the lightly doped
case, when the classical spin and orbital ordered ground state
present in the half-filled case is robust, the strong-coupling
model (\ref{eq:str-coupl}) can be reduced to an effective model
\begin{align}\label{eq:polaron}
{\cal H}={\cal H}_t+{\cal H}_J+{\cal H}_{3s},
\end{align}
see Eqs. (\ref{eq:hjlsw}) and
(\ref{eq:htpolaron})--(\ref{eq:h3spolaron}). Actually, this is a
polaron-type model with the coupling between fermions (holes) and
bosonic excitations (orbitons and magnons) which is relatively
easy to solve, see below. The validity of the mapping between the
two models was thoroughly discussed in Ref. \onlinecite{Mar91} and
\onlinecite{Woh08}.

Note, that the original strong-coupling model (\ref{eq:str-coupl})
has three parameters $\{J, \eta, t\}$, whereas the effective
polaron model given by Eq. (\ref{eq:polaron}) is more conveniently
analyzed using instead four parameters $\{J_S, J_O, \tau, t\}$,
which determine the scale of spin and orbital excitations, as well
as free hole propagation (due to the three-site terms) and the
vertex function ($t$), see below. In what follows we will use
either one of these two parameter sets (or even both of them),
depending on the context. Hence, we plot in Fig. \ref{fig:2} the
functional relation between the parameters $\{J_S,J_O,\tau\}$ on
Hund's exchange $\eta$. While the magnon energies $\propto J_S$
decrease with $\eta$, the energy scale of orbitons $\propto J_O$
increases rapidly, so the latter excitations are expected to play an
important role in the realistic regime of parameters.

\section{Green's functions and spectral functions }
\label{sec:results}

\subsection{The self-consistent Born approximation}
\label{sec:green's}

The spectral properties of the hole doped into the AF/AO ground
state $|\Phi_0\rangle$ with energy $E_0$ of the strong-coupling
model Eq. (\ref{eq:str-coupl}) at half-filling follow from the
Green's functions:
\begin{eqnarray}
G_{a \downarrow} ({\bf k}, \omega) &=& \left\langle\Phi_0 \left|
a^\dag_{\bf k \downarrow} \frac{1}{\omega+H-E_0} a_{\bf k
\downarrow} \right|\Phi_0\right\rangle\,,
\\
G_{a \uparrow} ({\bf k}, \omega) &=& \left\langle\Phi_0 \left|
a^\dag_{\bf k \uparrow} \frac{1}{\omega+H-E_0} a_{\bf k \uparrow}
\right|\Phi_0\right\rangle\,,
 \\
G_{b \downarrow} ({\bf k}, \omega) &=& \left\langle\Phi_0 \left|
b^\dag_{\bf k \downarrow} \frac{1}{\omega+H-E_0}
b_{\bf k \downarrow} \right|\Phi_0\right\rangle\,, \\
G_{b \uparrow} ({\bf k}, \omega) &=& \left\langle\Phi_0 \left|
b^\dag_{\bf k \uparrow} \frac{1}{\omega+H-E_0} b_{\bf k
\uparrow}\right|\Phi_0 \right\rangle\,.
\end{eqnarray}
However, due to the mapping of the strong-coupling model onto the
polaron model performed in Secs. \ref{sec:undoped}--\ref{sec:parameters},
it is now convenient to express the above Green's functions in
terms of the polaron Hamiltonian ${\cal H}$. This requires that we
first write down the electron operators in terms of the operators
used in Eq. (\ref{eq:polaron}):
\begin{align}
a_{\bf k \downarrow} &= \frac{1}{\sqrt{N}} \left(\sum_{{\bf j} \in A}
e^{i{\bf kj}} h^\dag_{\bf j}
+\sum_{{\bf j} \in B} e^{i{\bf kj}} h^\dag_{\bf j} \alpha^{}_{\bf j} \beta^{}_{\bf j} \right), \\
a_{\bf k \uparrow} &= \frac{1}{\sqrt{N}}
\left(\frac{1}{\sqrt{2}}\sum_{{\bf j} \in A} e^{i{\bf kj}} h^\dag_{\bf j}
\alpha^{}_{\bf j}
+\sum_{{\bf j} \in B} e^{i{\bf kj}} h^\dag_{\bf j} \beta^{}_{\bf j} \right), \\
b_{\bf k \downarrow} &= \frac{1}{\sqrt{N}} \left(\sum_{{\bf j} \in A}
e^{i{\bf kj}} h^\dag_{\bf j} \beta^{}_{\bf j}
+\frac{1}{\sqrt{2}}\sum_{{\bf j} \in B} e^{i{\bf kj}} h^\dag_{\bf j} \alpha^{}_{\bf j}\right), \\
b_{\bf k \uparrow} &= \frac{1}{\sqrt{N}} \left(\sum_{{\bf j} \in A}
e^{i{\bf kj}} h^\dag_{\bf j} \alpha^{}_{\bf j} \beta^{}_{\bf j} +\sum_{{\bf j} \in B}
e^{i{\bf kj}} h^\dag_{\bf j}  \right).
\end{align}
Second, the ground state $|\Phi_0\rangle$ is now a physical vacuum
$|0\rangle$, with respect to the Bogoliubov operators
$\tilde{\alpha}_{{\bf k}\pm}$ and the operators $\beta_{\bf k}$,
with energy $E$. Then, we arrive at the following relations:
\begin{align}
G_{a \downarrow} ({\bf k}, \omega) &=\frac{1}{2} \left\langle
0\left|h_{{\bf k}A}\frac{1}{\omega+{\cal H}-E}
h^\dag_{{\bf k} A} \right| 0 \right\rangle \nonumber \\
&\equiv \frac{1}{2}\;G_{AA}({\bf k}, \omega),  \\
G_{b \uparrow} ({\bf k}, \omega) &=\frac{1}{2} \left\langle 0
\left| h_{{\bf k} B} \frac{1}{\omega+{\cal H}-E}
h^\dag_{{\bf k} B} \right| 0 \right\rangle \nonumber \\
&\equiv \frac{1}{2}\;G_{BB}({\bf k}, \omega),
\end{align}
whereas
\begin{align}
G_{a \uparrow} ({\bf k}, \omega)\ll G_{a\downarrow}({\bf k},\omega)\,, \\
G_{b \downarrow}({\bf k},\omega)\ll G_{b\uparrow} ({\bf
k},\omega)\,,
\end{align}
and we skip the two latter Green's functions in what follows. Note
that the above set of equations follows from the fact that
$\beta_{\bf k} |0\rangle = 0$ and the inequalities are due to
\begin{align}
\left\langle 0 \left| \frac{\alpha_i^\dag\alpha_i}{2}\right|0
\right\rangle \sim \frac{n_0}{2} \sim 0.1\,,
\end{align}
where $n_0$ is the average number of spin deviations in the 2D
ground state $|0\rangle$. Note that below we will eliminate the
ground state energy $E$ to simplify equations.

\begin{figure}
\includegraphics[width=0.45\textwidth]{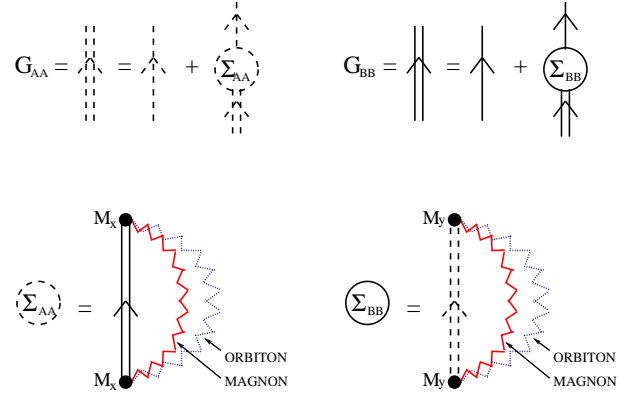}
\caption{(Color online) Diagrammatic representation of the SCBA
equations: top --- the Dyson's equation for the $G_{AA}({\bf k},
\omega)$ and $G_{BB}({\bf k}, \omega)$ Green's functions; bottom
--- equations for the respective self-energies. Note the
appearance of the two wiggly lines in the calculation of the
self-energies coming from simultaneous orbiton and the magnon
excitations. } \label{fig:3}
\end{figure}

As seen in Section \ref{sec:coupling} the vertices in the
spin-orbital model are slightly more complex than for the standard
spin $t$-$J$ model:\cite{Mar91} (i) one always finds two boson and
two holon lines at each vertex (instead of just one boson and two
holon lines), (ii) the two sublattice structure plays an essential
role (this resembles the orbital case), and (iii) one has two
kinds of magnons (which we had to introduce in order to keep track
of the lattice index). Also, in the LSW and LOW order all the
terms $\propto \alpha^\dag_{\bf i}\beta^{}_{\bf i} $ do not contribute (the
self-energies for them would require altogether four boson lines
instead of just two for $\propto\alpha^{}_{\bf i}\beta^{}_{\bf i}$).
\cite{note1} Hence, we largely follow the route proposed for the
orbital $t_{2g}$ model \cite{Woh08} and obtain the following SCBA
equations for the self-energies (see also Fig. \ref{fig:3}):
\begin{align} \label{eq:scba1}
&\Sigma_{AA} ({\bf k}, \omega) = 2\frac{z^2t^2}{4N^2} \sum_{\bf q_1, q_2 }
M^2_x({\bf k, q_1, q_2}) \nonumber \\
& \times G_{BB}({\bf k - q_1 -q_2}, \omega+J_S\omega_{\bf q_1}+J_0)\,, \\
&\Sigma_{BB} ({\bf k}, \omega) = 2\frac{z^2t^2}{4N^2} \sum_{\bf q_1, q_2 }
M^2_y({\bf k, q_1, q_2})  \nonumber \\
& \times G_{AA}({\bf k - q_1 -q_2}, \omega+J_S\omega_{\bf
q_1}+J_0)\,, \label{eq:scba2}
\end{align}
where the factor $2$ in front of each vertex comes from the fact
that we have two kinds of magnons [$+$ and $-$, see Eq. (\ref{eq:alpha})]
and hence two distinct diagrams. Fortunately, this factor cancels
with one of the $2$ in the denominator and we obtain that the
coupling constant is $(t/\sqrt{2})^2$, i.e., we recover this factor
$1/\sqrt{2}$ which originates from the quantum double exchange.\cite{Loo01}
The above equations should always be supplemented by the Dyson's equations:
(Fig. \ref{fig:3}),
\begin{align}\label{eq:dyson1}
G_{AA}({\bf k}, \omega)=\frac{1}{\omega+\tau \varepsilon_A({\bf k})
-\Sigma_{AA}({\bf k}, \omega)}\,, \\
G_{BB}({\bf k}, \omega)=\frac{1}{\omega+\tau \varepsilon_B({\bf
k})-\Sigma_{BB}({\bf k},\omega)}\,, \label{eq:dyson2}
\end{align}
which together form a self-consistent set of equations which can
be solved numerically by iteration.

Finally, we can calculate the spectral functions for a hole
created in $a$ and $b$ orbital:
\begin{eqnarray}
A_a({\bf k},\omega)&=&-\frac{2}{\pi}\lim_{\delta\to
0}\,\mbox{Im}\,G_{a \downarrow}({\bf k},\omega+i\delta) \nonumber
\\ &=& -\frac{1}{\pi}\lim_{\delta\to
0}\,\mbox{Im}\, G_{AA}({\bf k}, \omega+i\delta)\,, \\
A_b({\bf k},\omega)&=&-\frac{2}{\pi}\lim_{\delta\to
0}\,\mbox{Im}\,G_{b \uparrow}({\bf k},\omega+i\delta)\nonumber
\\&=&
-\frac{1}{\pi}\lim_{\delta\to 0}\,\mbox{Im}\, G_{BB}({\bf k},
\omega+i\delta)\,,
\end{eqnarray}
where we introduced a factor of $2$ in front of the definition of
the spectral function $A_\gamma({\bf k},\omega)$ for convenience.

\subsection{Analytic structure and quasiparticle pole}
\label{sec:analytic}

It occurs that in the case when the three-site terms are absent
(i.e., at $\tau \equiv 0$) one can easily prove two important
properties of the SCBA equations
(\ref{eq:scba1})--(\ref{eq:scba2}): (i) the self-energies are
${\bf k}$-independent, (ii) the spectral functions contain the
quasiparticle state for finite value of the exchange parameter $J$
(equivalently $J_S$ {\it or} $J_O$).

First, we show property (i). Since we assumed that $\tau = 0$, we
can rewrite SCBA equations (\ref{eq:scba1})-(\ref{eq:scba2})
together with Dyson's equations
(\ref{eq:dyson1})-(\ref{eq:dyson2}) in the following manner:

\begin{widetext}
\begin{eqnarray}
\label{eq:proof1}
\Sigma_{AA}({\bf k}, \omega)&=&\frac{z^2t^2}{2N^2} \sum_{\bf q_1,
q_2 } \frac{M^2_x ({\bf k, q_1, q_2})} {\omega+J_S\omega_{{\bf
q}_1}+J_O-\Sigma_{BB}({\bf k - q_1 -q_2}, \omega+J_S\omega_{\bf
q_1}+J_0)}\,, \\
\label{eq:proof2} \Sigma_{BB}({\bf k}, \omega)&=&
\frac{z^2t^2}{2N^2} \sum_{\bf q_1, q_2 } \frac{M^2_y ({\bf k, q_1,
q_2})} {\omega+J_S\omega_{{\bf q}_1}+J_O-\Sigma_{AA}({\bf k - q_1
-q_2}, \omega+J_S\omega_{\bf q_1}+J_0)}\,,
\end{eqnarray}
which after substitution ${\bf q}_2 \rightarrow {\bf k}-{\bf q}_2$
in the sums leads to
\begin{eqnarray}\label{eq:proof3}
\Sigma_{AA} ({\bf k},\omega)&=&\frac{z^2t^2}{2N^2} \sum_{\bf q_1,
q_2 } \frac{f^2({\bf q}_1, {\bf q}_2)} {\omega+J_S\omega_{{\bf
q}_1}+J_O-\Sigma_{BB}({\bf q_2 -q_1},
\omega-J_S\omega_{\bf q_1}-J_0)}\,, \\
\Sigma_{BB} ({\bf k}, \omega)&=&\frac{z^2t^2}{2N^2} \sum_{\bf q_1,
q_2 } \frac{g^2({\bf q}_1, {\bf q}_2)} {\omega+J_S\omega_{{\bf
q}_1}+J_O-\Sigma_{AA}({\bf q_2 -q_1}, \omega+J_S\omega_{\bf
q_1}+J_0)}\,, \label{eq:proof4}
\end{eqnarray}
\end{widetext}

\noindent where we have used a short-hand notation:
\begin{eqnarray}
f({\bf q}_1, {\bf q}_2)&=&u_{{\bf q}_1} \gamma_{q_{2x}-q_{1x}} +
v_{{\bf q}_1} \gamma_{q_{2x}}\,, \\
g({\bf q}_1, {\bf q}_2)&=&u_{{\bf q}_1} \gamma_{q_{2y}-q_{1y}} +
v_{{\bf q}_1} \gamma_{q_{2y}}\,.
\end{eqnarray}
Since the self-energies on the right hand side of Eqs.
(\ref{eq:proof3})--(\ref{eq:proof4}) do not depend on ${\bf k}$ we
are allowed to drop the momentum dependence of the self-energies.
Hence, the spectral functions are also momentum independent in the
case of $\tau = 0$ and we recognize that the dependence of the
spectral functions on ${\bf k}$ may originate only from the
three-site terms.

Second, using the dominant pole approximation\cite{Kan89} we show
that the quasiparticle state exists [property (ii)] if $J$ is
finite (i.e. $J_S$ {\it or} $J_O$ are finite). Hence, following
Kane {\it et al.},\cite{Kan89} we assume that the Green's function
can be separated into the part containing the pole and the part
responsible for the incoherent processes:
\begin{eqnarray}
\label{eq:pole1}
G_{AA}(\omega)&=&\frac{a_A}{\omega-\omega_A}+G^{inc}_{AA}(\omega)\,, \\
\label{eq:pole2}
G_{BB}(\omega)&=&\frac{a_B}{\omega-\omega_B}+G^{inc}_{BB}(\omega)\,,
\end{eqnarray}
where
\begin{align}
a_A=\frac{1}{1-\frac{\partial \Sigma_{AA}(\omega)}{\partial \omega}
\big|_{\omega=\omega_A}}\,, \\
a_B=\frac{1}{1-\frac{\partial \Sigma_{BB}(\omega)}{\partial \omega}
\big|_{\omega=\omega_B}}\,,
\end{align}
and the pole positions:
\begin{eqnarray}
\omega_A&=&\Sigma_{AA}(\omega_A)\,, \\
\omega_B&=&\Sigma_{AA}(\omega_B)\,,
\end{eqnarray}
have to be determined self-consistently following Eqs.
(\ref{eq:pole1})--(\ref{eq:pole2}).

Next, following Ref. \onlinecite{Kan89}, it is straightforward to
derive the upper bound for the residues $\{a_A,a_B\}$:
\begin{eqnarray}\label{eq:aA}
a_A\!&\leq& \left\{1+\frac{z^2t^2}{2N^2}\sum_{{\bf q_1,
q_2}}\!f^2({\bf q}_1, {\bf q}_2) \frac{a_B}{(J_S\omega_{{\bf
q}_1}+J_0)^2}\right\}^{-1}\,, \nonumber \\ \\
\label{eq:aB} a_B\!&\leq& \left\{1+\frac{z^2t^2}{2N^2}\sum_{{\bf
q_1, q_2}}\!g^2({\bf q}_1, {\bf q}_2) \frac{a_A}{(J_S\omega_{{\bf
q}_1}+J_0)^2}\right\}^{-1}\,. \nonumber \\
\end{eqnarray}
If the sums in the above equations are divergent, then the upper
bounds for the residues are equal zero and the Green's functions
do not have the quasiparticle pole.  Hence, we need to check the
behavior for small values of the momenta ${\bf q}_1$. Then
$\omega_{{\bf q}_1}\propto |{\bf q}_1|$, but e.g. $(u_{{\bf q}_1}
\gamma_{q_{2y}-q_{1y}} + v_{{\bf q}_1} \gamma_{q_{2y}} )^2 \sim
|{\bf q}_1|(\gamma_{q_{2y}}-\widehat{q}_1 \cdot \nabla
\gamma_{q_{2y}})^2$. Thus, if at least either $J_S$ or $J_O$ is
finite, then there are no divergences in Eqs.
(\ref{eq:aA})--(\ref{eq:aB}). Consequently, under the same
conditions the quasiparticle state exists.

\section{NUMERICAL RESULTS}
\label{sec:num}

\subsection{Spectral functions}
\label{sec:spectral}

\begin{figure}[t!]
\includegraphics[width=0.45\textwidth]{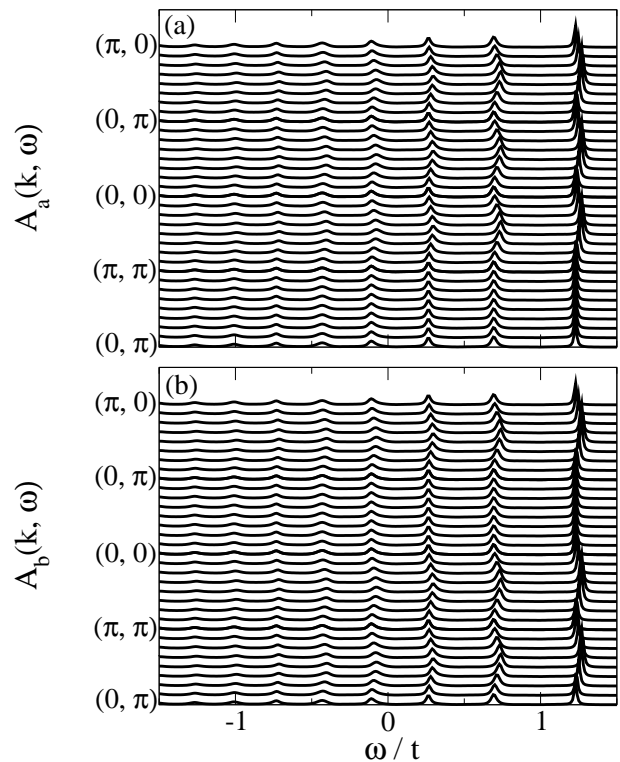}
\caption{The spectral functions derived within the SCBA for the
hole doped at half-filling to the spin-orbital model
(\ref{eq:str-coupl}) into: (a) $a$ orbital, and (b) $b$ orbital.
Parameters: $J=0.2t$ and $\eta=0.15$ (i.e., $J_S=0.03t$,
$J_O=0.08t$, and $\tau=0.06t$). Broadening $\delta=0.01t$ and
cluster size $16\times16$.} \label{fig:4}
\end{figure}

\begin{figure}[t!]
\includegraphics[width=0.45\textwidth]{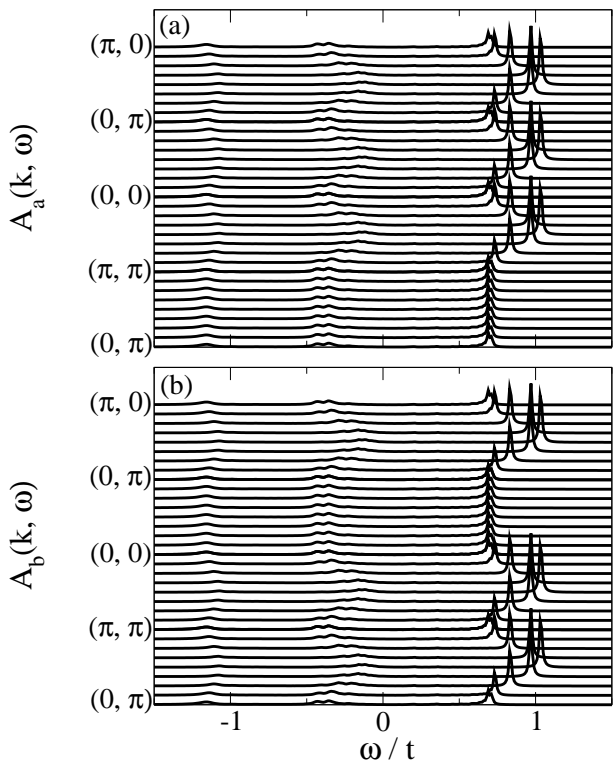}
\caption{The spectral functions derived within the SCBA for the
hole doped at half-filling to the spin-orbital model
(\ref{eq:str-coupl}) into: (a) $a$ orbital, and (b) $b$ orbital.
Parameters: $J=0.6t$ and $\eta=0.15$ (i.e., $J_S=0.09t$,
$J_O=0.23t$, and $\tau=0.19t$). Broadening $\delta=0.01t$ and
cluster size $16\times16$.} \label{fig:5}
\end{figure}

We calculated the spectral functions $A_a({\bf k},\omega)$ and
$A_b({\bf k},\omega)$ by solving SCBA equations
(\ref{eq:scba1})--(\ref{eq:scba2}) on a mesh of $16\times16$ ${\bf
k}$-points. The results for two different values of the
superexchange constant $J=0.2t$ and $J=0.6t$ are shown in Figs.
\ref{fig:4} and \ref{fig:5}, respectively. Also we assume the
Hund's coupling to be close to a realistic value in LaVO$_3$,
i.e., $\eta=0.15$.\cite{note2} We emphasize that such a value of
the Hund's coupling is not only realistic but together with the
observed value of $J=0.2t$ gives a reasonable value of the
spin-only exchange constant $J_S=0.03t$ (which is in agreement
with the observed value of the N\'eel temperature in LaVO$_3$
$T_N\simeq 143$ K),\cite{Miy00,Miy06} cf. Sec. VIB of Ref.
\onlinecite{Ole05} for a more detailed discussion on this issue.
Thus, the spectral functions shown in Fig. \ref{fig:4} are
calculated for the realistic values of parameters in LaVO$_3$
whereas those shown in Fig. \ref{fig:5} will serve as a
comparison with the regime of large $J$.

Let us first discuss the results presented in Fig. \ref{fig:4}.
The quasiparticle peak in the low energy part of the spectrum is
clearly visible and confirms the analytic predictions presented in
Sec. \ref{sec:analytic}. However, the quasiparticle state has a
rather weak 1D dispersion: along the $k_x$ direction for holes
doped into the $b$ orbitals and along the $k_y$ direction for
holes doped into the $a$ orbital. Since such a dispersion was not
present when the three-site terms were neglected (see Sec.
\ref{sec:analytic}) we can immediately ascribe the onset of the 1D
dispersion in the spectra to the three-site terms. Indeed, these
terms are responsible for the coherent hole propagation in the
ground state with AO order and the hole dispersion is dressed by
incoherent processes. This phenomenon is quite well understood in
the framework of the orbital-only strong-coupling model
--- we refer to Ref. \onlinecite{Dag08} and Ref.
\onlinecite{Woh08} for the more detailed discussion.

Besides, the excited states form a ladder-like spectrum, also with
a small 1D dispersion. Similarly to the quasiparticle state
the excited states resemble qualitatively the spectral functions
calculated for the purely orbital model, see Ref.
\onlinecite{Dag08} and Ref. \onlinecite{Woh08}. However, here we
see that the spin-orbital spectra are quantitatively different
than the orbital ones. In particular, they are quite similar to
those obtained for $J=0.4t$ in the purely orbital model (see Fig.
9 in Ref. \onlinecite{Woh08}) although this is just a coincidence
since all of the exchange constants used in the spin-orbital model
($J$, $J_S$ or $J_O$) are much smaller than $0.4t$.

Looking at the spectral functions shown in Fig. \ref{fig:5} we
note that the ladder-like spectrum almost disappears and the
quasiparticle dispersion increases to $\sim 0.4t$ when the value
of the superexchange constant is increased to $J=0.6t$. However,
the qualitative features in the dispersion of the quasiparticle
peak stay mostly unchanged and the 1D character of the dispersion
relation is preserved. Quantitatively, both the quasiparticle
spectral weight and the bandwidth increase quite drastically.

To conclude, the spin-orbital spectral functions form ladder-like
spectra with a small 1D dispersion and have many similarities with
the purely orbital spectra of the $t_{2g}$ model. \cite{Dag08}
Still, however, there are few relatively important differences
with the orbital model which suggest that the spin-orbital case is more
complex and the quasiparticle behavior more subtle than for the
purely orbital model. For the understanding of this problem we
refer to Sec. \ref{sec:discussion} whereas in the next section we
further investigate the properties of the quasiparticle arising in
the spectral functions of the spin-orbital model.

\subsection{Quasiparticle properties}
\label{sec:quasiparticle}

\begin{figure}[t!]
\includegraphics[width=0.35\textwidth]{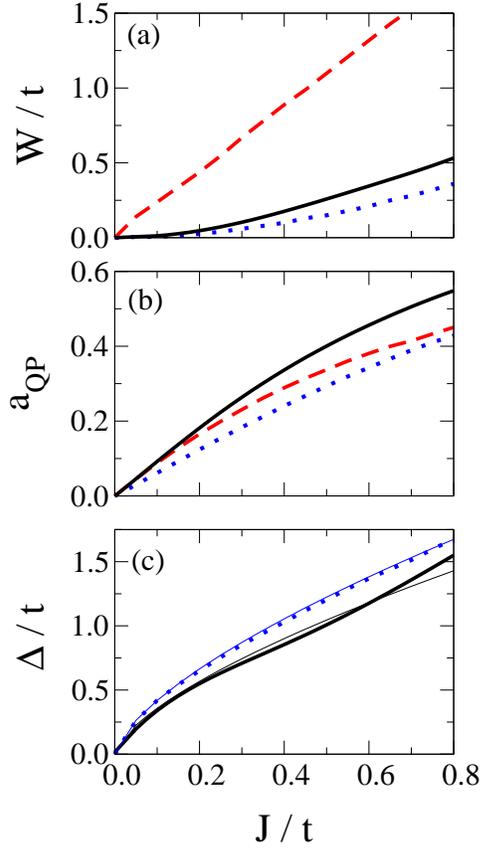}
\caption{(Color online) Quasiparticle properties for increasing
$J$ for the spin-orbital model discussed in the present paper with
the Hund's coupling $\eta=0.15$ (solid line), compared with the
spin model of Ref. \onlinecite{Bal95} (dashed line), and with the
orbital model of Ref. \onlinecite{Dag08} (dotted line). Different
panels show: (a) the quasiparticle bandwidth $W$, (b) the
quasiparticle spectral weight $a_{\rm QP}$ averaged over the 2D
Brillouin zone, and (c) the pseudogap $\Delta$ (energy distance between the
quasiparticle state and the first excited state) at ${\bf
k}=(\pi/2,\pi/2)$. Note that $\Delta$ is not shown for the spin
model as it cannot be defined there. The two light solid lines on
panel (c) show the $(J/t)^{2/3}$ curves fitted to the data at low
values of $J$ for the orbital and the spin-orbital model.}
\label{fig:6}
\end{figure}

In this section we analyze in more detail the quasiparticle
 properties of the spectral functions of the
spin-orbital model, cf. Fig. \ref{fig:6}. In addition, we compare
these properties with those calculated for the spin
strong-coupling model of Ref. \onlinecite{Bal95} and the orbital
strong-coupling model of Ref. \onlinecite{Dag08}.

By looking at the results for the spin-orbital model in Fig.
\ref{fig:6} we see that: (i) the quasiparticle bandwidth $W$ is
strongly renormalized from its respective free value (which is $W=8 \tau$
as it may only originate from the three-site term dispersion
relation, see Secs. \ref{sec:analytic}-\ref{sec:spectral}) and is
proportional to $J^2$ for small $J$ ($J\lesssim 0.6t$) and to $J$ in the
regime of large $J$ ($J\gtrsim 0.6t$); (ii) the quasiparticle spectral
weight $a_{\rm QP}$ grows considerably with increasing $J$; (iii)
the pseudogap $\Delta$ (the energy distance between the
quasiparticle state and the first excited state) exists and
roughly scales as $t (J/t)^{2/3}$ although for larger values of
superexchange $J$ some deviations from this law were observed.

Thus, on the one hand, we see that the quasiparticle properties of
the spin-orbital model are qualitatively distinct than those of
the spin model. We should stress that although the spin model used
here is the $S=1/2$ $t$-$J$ model with the three-site
terms\cite{Bal95} but we expect similar generic behavior in the
case of spin $S=1$ (see also Sec. \ref{sec:spin}). Hence, as there
is no qualitative agreement with the model for spin $S=1/2$ there
would neither be a qualitative agreement with the model for spin
$S=1$.

On the other hand, we note remarkable similarities between the
present model and the orbital model (see also Fig. \ref{fig:6});
the generic behavior of the bandwidth, the spectral weight, and of
the pseudogap seems to be almost identical in both models. There
are, however, two important differences. First, the $t(J/t)^{2/3}$
law does not describe the behavior of the spin-orbital pseudogap
so well as in the orbital case. Second, if we assume that the
spin-orbital model is just qualitatively similar to the orbital
model, then it should be possible to rescale all the quasiparticle
properties with such an effective value of the superexchange $J$
that they coincide with the results for the orbital model. This,
however, is not possible. For example, from the fits to the $t
(J/t)^{2/3}$ law we can deduce that such an effective value of the
superexchange would be $J_{\rm eff}=(a_{SO}/a_{O})^{2/3}=0.79J$,
where $a_{SO}=1.94$ ($a_{O}=1.66$) is the fitted coefficient which
multiplies the $t(J/t)^{2/3}$ law in the spin-orbital (orbital)
case. This single parameter does not suffice, however, as a
similar fit to the quasiparticle spectral weight would require
such a $J_{\rm eff}$ to be bigger than $J$.

In conclusion, the properties of the quasiparticle state in the
spin-orbital model resemble those found for the quasiparticle in
the purely orbital $t_{2g}$ model.\cite{Dag08, Woh08} However, a more
detailed analysis presented in this section reveals that such a
correspondence is rather "superficial" and that few of the
qualitative features of both models are different. We refer to the
next section for the thorough understanding of this phenomenon.

\section{ORIGIN OF SPIN-ORBITAL POLARONS}
\label{sec:discussion}

\subsection{Comparison with spin and orbital polarons}
\label{sec:spin}

The main task of this and the following two sections is to understand
the spectral functions and the quasiparticle properties of the
spin-orbital model. In particular, we not only want to understand
the peculiar similarities or rather small differences between
the spin-orbital model and its purely orbital counterpart, which
were discussed in the last two sections (see Secs.
\ref{sec:spectral} and \ref{sec:quasiparticle}), but also we want
to understand why the spin physics seems to be "hidden" in the
present spin-orbital system. Note that for the sake of simplicity
in this section we will entirely neglect the three-site terms
(\ref{eq:h3s}) in the spin-orbital model (\ref{eq:str-coupl}) (or
in other words we will put $\tau\equiv0$). This is motivated by
the fact that the role of the three-site terms is solely to
provide dispersion to the spectra (see Secs. \ref{sec:analytic}-
\ref{sec:quasiparticle}) and the origin and character of these
dispersive features is quite well understood by now, see. Ref.
\onlinecite{Woh08}.

We start the analysis of spectral functions by introducing two
artificial toy models in order to extract the features related to
spin and orbital excitations separately, whose results will be
later compared to those of the $t$-$J$ spin-orbital model. First,
we define the following $t$-$J_S$ toy-spin model
\begin{equation}
\label{eq:spinmodel} H_{S}\!=-t \sum_{\langle {\bf i}{\bf j}\rangle, \sigma}
{\cal P}( \tilde{c}^\dag_{{\bf i}\sigma} \tilde{c}_{{\bf j}\sigma} + {\rm
H.c.}){\cal P}+ J_S \sum_{\langle {\bf i}{\bf j}\rangle} {\bf S}_{\bf i}\cdot{\bf
S}_{\bf j}\,,
\end{equation}
where spin operators $\{{\bf S}_{\bf i}\}$ stand for $S=1$ spins, the
constrained operators,
$\tilde{c}^\dag_{{\bf i}\sigma}=c^\dag_{{\bf i}\sigma}(1-n_{{\bf i}\bar{\sigma}})$,
exclude double occupancies from the Hilbert space, and the
operators ${\cal P}$ project onto the high spin states. Note that
here the superexchange energy scale is $J_S$ [see Eq.
(\ref{eq:js})] and not $J$. Hence, it is defined in such a way
that it mimics the influence of AO order on the spin subsystem. On
the other hand, the kinetic energy is blind to the AO here and Eq.
(\ref{eq:ht}) reduces to the kinetic part of Eq.
(\ref{eq:spinmodel}) only if the orbitals form an orbital liquid
state. This is an obvious logical inconsistency but our aim here
is to see what happens when the joint spin-orbital dynamics in the
kinetic energy is entirely neglected. It is also the reason why we
call model (\ref{eq:spinmodel}) the toy-spin model.

\begin{figure}[t!]
\includegraphics[width=8.2cm]{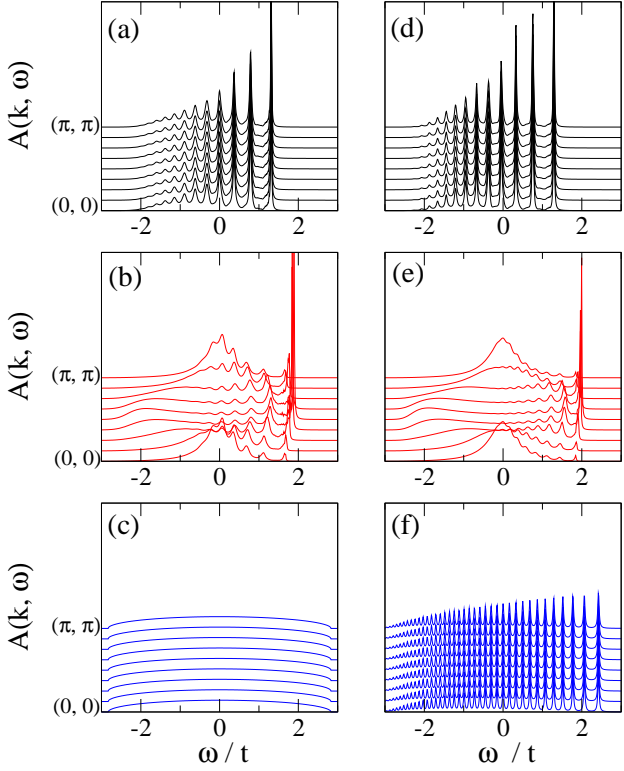}
\caption{(Color online) Spectral function along the $\Gamma-M$
direction of the Brillouin zone for: the spin-orbital model (top
panels), toy-spin model (middle panels), toy-orbital model (bottom
panels). Parameters: $J=0.2t$, $\eta=0$ (i.e., $J_S=0.05t$,
$J_O=0$) on the left panels whereas $J=0.2t$, $\eta=0.15$ (i.e.,
$J_S=0.03t$, $J_O=0.08t$) on the right panels; in both cases
$\tau\equiv 0$, see text. Note that $\tau\equiv 0$ implies
\cite{Woh08} $A_a({\bf k}, \omega)= A_b({\bf k},\omega) \equiv
A({\bf k},\omega)$. Broadening $\delta=0.01t$ [$\delta=0.02t$ on
panel (f)] and cluster size $16\times16$.} \label{fig:7}
\end{figure}

Second, we define the following $t$-$J_O$ toy-orbital model
\begin{equation}
\label{eq:orbitalmodel} H_{O}=-t \sum_{\bf i}\left(
\tilde{b}^\dag_{\bf i} \tilde{b}_{{\bf i}+\hat{\bf a}}
+\tilde{a}^\dag_{{\bf i}}\tilde{a}_{{\bf i}+\hat{\bf b}} + {\rm H.c.}\right)
+ J_O \sum_{\langle {\bf i}{\bf j} \rangle}T^z_{\bf i} T^z_{\bf j},
\end{equation}
where pseudospin $T=1/2$, and the constrained operators
$\tilde{b}^\dag_{\bf i}=b^\dag_{\bf i}(1-n_{{\bf i}a})$ and
$\tilde{a}^\dag_{\bf i}=a^\dag_{\bf i}(1-n_{{\bf i}b})$ exclude double occupancies.
Similarly as in the toy-spin model defined above, the
superexchange energy scale is not $J$ but $J_O$ [see Eq.
(\ref{eq:jo})] which mimics the influence of spin system with the
AF order on the orbital one. Also, the kinetic energy is blind now
to the AF order and Eq. (\ref{eq:ht}) reduces to the kinetic part
of Eq. (\ref{eq:orbitalmodel}) only if the spins form the FM
order. This again is logically inconsistent, but we wish to
consider this situation for the same reasons as for the spins, see
above.

\begin{figure}[t!]
\includegraphics[width=8.2cm]{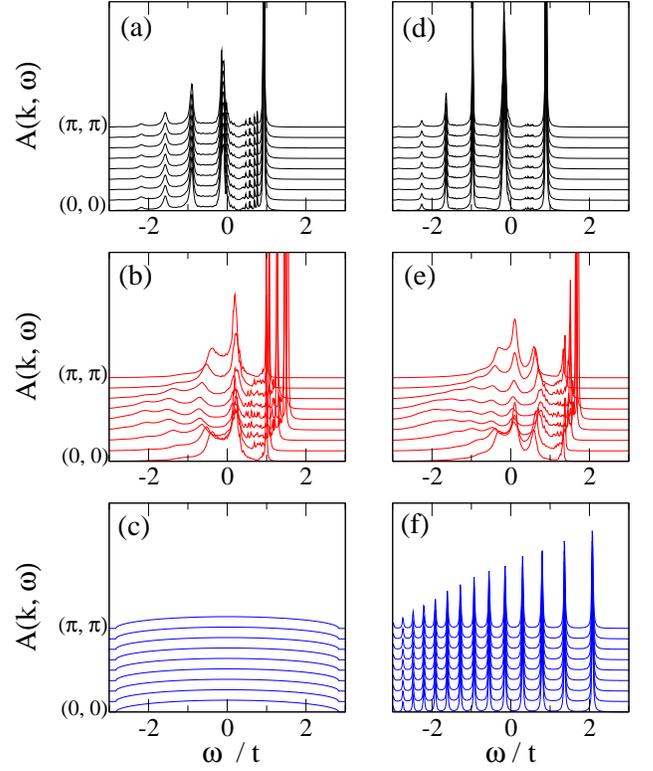}
\caption{(Color online) Spectral function along the $\Gamma-M$
direction of the Brillouin zone for: the spin-orbital model (top
panels), toy-spin model (middle panels), toy-orbital model (bottom
panels). Parameters: $J=0.6t$, $\eta=0$ (i.e., $J_S=0.15t$,
$J_O=0$,) on the left panels, whereas $J=0.6t$, $\eta=0.15$ (i.e.,
$J_S=0.09t$, $J_O=0.23t$) on the right panels; in both cases
$\tau\equiv 0$, see text. Note that $\tau\equiv 0$ implies
\cite{Woh08} $A_a({\bf k}, \omega)= A_b({\bf k},\omega) \equiv
A({\bf k},\omega)$. Broadening $\delta=0.01t$ [$\delta=0.02t$ on
panel (f)] and cluster size $16\times16$. } \label{fig:8}
\end{figure}

Next, we solve these models using the SCBA method in the case of
the one hole doped into the AF (AO) state for the toy-spin
(toy-orbital) model. We do not show here the respective SCBA
equations as these are straightforward and follow from those
written in Ref. \onlinecite{Mar91} (Ref. \onlinecite{Woh08}) in
the case of the spin (orbital) model. One only has to remember to
substitute in the respective SCBA equations $J\rightarrow -J_S$,
$S\rightarrow 1$ and (due to the quantum double exchange factor) also
$t\rightarrow t/\sqrt{2}$ in Ref. \onlinecite{Mar91} in the spin
case, and $J\rightarrow -J_O$, $E_0\rightarrow 0$, and
$\tau\rightarrow 0$ in Ref. \onlinecite{Woh08} in the orbital
case. We then calculate the respective spectral functions on a
mesh of $16\times16$ ${\bf k}$-points. Note that since $\tau\equiv
0$ the spectral functions for both orbital flavors are equal, i.e.
$A_a({\bf k},\omega)=A_b({\bf k},\omega)\equiv A({\bf k},\omega)$.

Finally, we compare the results obtained for the above toy models
with those obtained for the spin-orbital $t$-$J$ model, Eq.
(\ref{eq:ht}) and (\ref{eq:hj}). We show the results for two different
values of $J=0.2t$ and $0.6t$, see Figs. \ref{fig:7} and
\ref{fig:8}. In addition, we calculate the results for two
different values of the Hund's coupling $\eta=0$ and $\eta=0.15$
(see left and right panels of each figure).

Let us first look at the physical regime of finite $\eta=0.15$ and
$J=0.2t$, see Figs. \ref{fig:7}(d)--\ref{fig:7}(f). On
the one hand, the spin-orbital spectral function [panel (d)]
resembles qualitatively the ladder spectrum found in the orbital
model [panel (f)] although the quantitative comparison reveals
strong differences between the two models. On the other hand, the
spin-orbital spectral function is entirely different from the
${\bf k}$-dependent spin spectral function [panel (e)]. One also
finds a somewhat similar behavior for the case of $\eta=0.15$ and
$J=0.6t$, see Figs. \ref{fig:8}(d)--\ref{fig:8}(f). Here, however,
the spin-orbital spectrum is much more different than the orbital
spectrum, so the renormalization by the spin part is stronger.

Even more inquiring behavior is found in the nonphysical regime of
$\eta=0$ (which, however, is an interesting mathematical limit).
\cite{note2} Then neither of the panels shown in Figs.
\ref{fig:7}(a)--\ref{fig:7}(c) or Figs.
\ref{fig:8}(a)--\ref{fig:8}(c) is similar to each other. This
means that even the orbital model is entirely different in this
regime than the spin-orbital model. This is because in this limit
the hole moves in the orbital model incoherently as $J_O=0$ for
$\eta=0$, see e.g. Fig. \ref{fig:7}(c). However, apparently in the
spin-orbital model with $J_O=0$ and small but finite $J_S$ the
hole moves in string-like potential,  see e.g. Fig.
\ref{fig:7}(a). This means that the onset of the ladder-like
spectrum in the spin-orbital model in this regime cannot be
explained easily in terms of the purely orbital model.

In addition, one sees that whereas for different values of $\eta$
but the same values of $J$ one gets rather similar spin-orbital
spectra, the spectra found for the toy-orbital model are
different. On the contrary, for different values of $J$ and the same
values of $\eta$ the spin-orbital spectra are rather distinct but
the spin spectra do not change much. This is another argument
which suggests that neither the toy-spin model nor the toy-orbital
model can explain alone the features found in the spin-orbital
spectra.

To conclude, we note that the joint spin-orbital dynamics in the
kinetic part of the spin-orbital model plays a significant role
for the coherent hole motion. When spin and orbital degrees of
freedom are separated from each other (disentangled), the purely
spin or orbital models cannot reproduce the spectral function
found for the spin-orbital $t$-$J$ model. Hence, indeed the
spin-orbital spectral functions resemble the orbital ones only
superficially and it is the peculiar interplay of the spins and
orbitals, studied in the next section, which leads to the
calculated spectra.

\subsection{Spin-only string excitations at $\eta=0$}
\label{sec:understanding}

In this section we attempt\cite{note2} to understand the
spin-orbital spectra in the limit of $\eta=0$ by assuming that the
spins $S=1$ are purely classical and interact by Ising
superexchange. Hence, we skip the spin-flip terms $\propto
S^+_{\bf i}S^-_{\bf j}$ in Hamiltonian (\ref{eq:hj}), rewrite SCBA
equations (\ref{eq:scba1})--(\ref{eq:scba2}) in this case, and
finally try to compare the spectral functions calculated in this
regime with the ones obtained for the full model, given by Eq.
(\ref{eq:str-coupl}). In addition, we not only assume $\eta=0$
(which implies $J_O=0$ in particular) but also $\tau\equiv0$
(similarly as in the previous section). In the next section we
discuss the impact of the finite value of these parameters on the
results obtained under these assumptions.

\begin{figure}[t!]
\includegraphics[width=7cm]{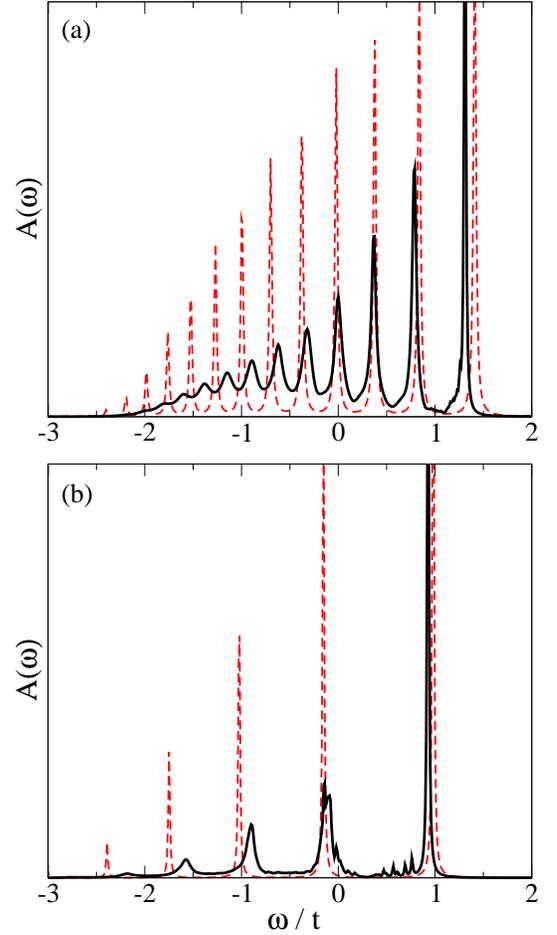}
\caption{(Color online) Comparison between the spectral functions
of the classical limit (dashed lines) and full quantum version
(solid lines) of the spin-orbital model at a single point in the
Brillouin zone (the spectra are ${\bf k}$-independent) for: (a)
$J=0.2t$ and $\eta=0$ (i.e., $J_S=0.05t$, $J_O=0$), and (b)
$J=0.6t$ and $\eta=0$ (i.e., $J_S=0.15t$, $J_O=0$); in both cases
$\tau\equiv 0$, see text. Note that $\tau\equiv 0$ implies
\cite{Woh08} $A_a(\omega)= A_b(\omega) \equiv A(\omega)$.
Broadening $\delta=0.01t$ ($\delta=0.015t$) in panel (a) [(b)] and
cluster size $16\times16$.} \label{fig:9}
\end{figure}

Since $u_{{\bf q}_1}=1$, $v_{{\bf q}_1}=0$, and $\omega_{{\bf
q}_1}=zS$ for the Ising superexchange,
\cite{Mar91} we can rewrite self-energy equations
(\ref{eq:proof1})--(\ref{eq:proof2}):
\begin{align}\label{eq:scbaclassical1}
&\Sigma_{AA} ({\bf k}, \omega) =\!\frac{z^2t^2}{2N^2}\!\sum_{\bf q_1, q_2 }
\frac{\gamma^2_{q_{1x}}}{\omega+J_S zS-\Sigma_{BB}({\bf q}_1, \omega+J_S zS)}\,, \\
&\Sigma_{BB} ({\bf k}, \omega) = \!\frac{z^2t^2}{2N^2}\!\sum_{\bf q_1,
q_2 } \frac{\gamma^2_{q_{1y}}}{\omega+J_S zS-\Sigma_{AA}({\bf q}_1,
\omega+J_S zS)}\,, \label{eq:scbaclassical2}
\end{align}
where we already substituted ${\bf q}_1\rightarrow {\bf k}-{\bf
q}_1-{\bf q}_2 $ in the sums. Then the above self-consistent
equations are momentum independent and we obtain:
\begin{align}\label{eq:scbaclassical3}
&\Sigma_{AA} ( \omega) ={\Big(\frac{t}{\sqrt{2}}\Big)}^2
\frac{z/2}{\omega+J_S zS-\Sigma_{BB}(\omega+J_S zS)}\,, \\
&\Sigma_{BB} ( \omega) ={\Big(\frac{t}{\sqrt{2}}\Big)}^2
\frac{z/2}{\omega+J_S zS-\Sigma_{AA}(\omega+J_S zS)}\,,
\label{eq:scbaclassical4}
\end{align}
since $1/N \sum_{{\bf q}_1}\gamma^2_{q_{1\nu}} =1/(2z)$, where
$\nu=x,y$.

We have solved Eqs.
(\ref{eq:scbaclassical3})--(\ref{eq:scbaclassical4})
self-consistently. The respective spectral functions are shown in
Fig. \ref{fig:9} for $J=0.2t$ (i.e., $J_S=0.05t$) and $J=0.6t$
(i.e., $J_S=0.15t$). As expected, one obtains a typical ladder
spectrum in the entire regime of $J$. However, we see that the
results resemble those obtained for the full spin-orbital model
(\ref{eq:str-coupl}) with $\eta=0$ and $\tau=0$: Although the
spectrum of the full spin-orbital model contains some incoherent
part, the ladder peaks of the full spin-orbital model and of its
classical version (with Ising spin superexchange) almost coincide.
In addition, the incoherent bandwidth in the $J=0$ limit is $W=4t$
in the classical case whereas it is only slightly reduced in the
quantum model ($W\sim3.7t$).\cite{note3} For finite $J$ this
results in the small shift of the peaks in the full spin-orbital
model with respect to its classical counterpart. Altogether, this
suggests that the classical and the full (quantum) versions of the
spin-orbital models are to a large extent equivalent.

We note that Eqs.
(\ref{eq:scbaclassical3})--(\ref{eq:scbaclassical4}) are almost
identical to the SCBA equations for the hole moving in the $S=1/2$
spin Ising model [cf. Eq. (20) in Ref. \onlinecite{Mar91}]. The
only differences are: the self-consistent dependence of the
self-energies on two different sublattices, the reduction of the
nearest neighbors by a factor $1/2$ (in the numerator), the
reduction factor $1/\sqrt{2}$ for the hopping, and the increase of
the magnon excitation energy by a factor of $2$. Whereas the first
two imply the zig-zag hole motion in the ordered state,
\cite{Woh08} the two others merely mean that the hole moves in the
spin $S=1$ system, cf. similar factors for the triplet hole in the
$t$-$J$ model.\cite{Zaa92} This demonstrates that the hole motion
in the full spin-orbital model with $\eta=0$ and $\tau=0$ can be
quite well approximated by the zig-zag hole motion in the $S=1$
spin Ising model.

The whole analysis presented above leads us to the conclusion that in
the limit of $J_0=0$ and $\tau=0$ the hole moves in the spin and
orbitally ordered plane in the following way: (i) the orbitals
force the hole to move along the zig-zag paths even in the limit
of $J_0=0$, (ii) the orbitals force the hole to retrace its path
again even in the limit of $J_0\to 0$ --- this resembles the
situation discussed by Brinkman and Rice \cite{Bri70} for the spin
system, where the hole in the Ising model with $J=0$ always has to
retrace its path, (iii) the coherent hole motion by the coupling
to the spin fluctuations cannot occur in the ground state as then
the hole would not retrace its path, so (iv) instead the {\it hole
creates strings in the spin sector} which are erased when the hole
retraces its path. Thus, we note here a complex interplay of spins
and orbitals. In particular, due to point (ii) {\it the orbitals force
the spins to effectively act on the hole as the classical objects}.

\subsection{Composite string excitations at $\eta>0$}
\label{sec:string}

One may wonder how to extend the above understanding of the role
of spin and orbital excitations in the formation of spin-orbital
polarons to the case of finite values of orbital exchange
interaction $J_O>0$ at $\eta>0$ (see Fig. \ref{fig:2}), or in the
presence of three-site terms at finite $\tau$. On the one hand,
including the nonzero value of $J_O$ merely leads to the
substitution of $J_S zS\rightarrow J_S zS+J_O$ in Eqs.
(\ref{eq:scbaclassical1})--(\ref{eq:scbaclassical2}) and
consequently also in Eqs.
(\ref{eq:scbaclassical3})--(\ref{eq:scbaclassical4}); this
increases the string-like potential acting on the hole, coming in
this case also from the orbital sector. On the other hand,
including the three-site term $\tau>0$ results in the energy shift
$\omega\rightarrow\omega+\varepsilon_A({\bf k})$ and
$\omega\rightarrow\omega+\varepsilon_B({\bf k})$ in Eqs.
(\ref{eq:scbaclassical1})--(\ref{eq:scbaclassical2}) which results
in the ${\bf k}$-dependence of the spectral functions, see Eqs.
(\ref{eq:scbaclassical3})--(\ref{eq:scbaclassical4}). However, we
can still solve the model using the SCBA. The results (not shown)
resemble those found in Figs. \ref{fig:4} and \ref{fig:5}: it is
again only the incoherent part which is slightly enhanced in the
full spin-orbital model (\ref{eq:str-coupl}) whereas in its
classical counterpart it is suppressed. Furthermore, all of the
quasiparticle properties of the full spin-orbital model shown in
Fig. \ref{fig:6} are {\it almost perfectly\/} reproduced by the
classical spin-orbital model (not shown) --- with the only slight
discrepancy occurring in the region of the slight deviation from
the $t(J/t)^{2/3}$ law in the pseudogap of the full spin-orbital
model.

Finally, there remains just one subtle issue: how to explain the
appearance of the small momentum independent incoherent part in
the high-energy part of the spectrum for the quantum spin-orbital
model, cf. Fig. \ref{fig:9} in the case of $\eta=0$ [a similar
incoherent dome is visible in the spectrum for finite $\eta$ (not
shown)]. This can be understood in the following way: although the
hole has to return to the original site (due to the orbitals) the
magnons present in the excited states can travel freely in the
system. Hence, the energies of the excited states can no longer be
classified merely by the length of the retraceable paths (as it
would be the case in the classical model with no dispersion in the
magnon spectrum). This results in the small incoherent spectrum
which surrounds each of the peak of the ladder spectrum visible in
for example Fig. \ref{fig:9}. Furthermore, this incoherent
spectrum grows with increasing $J_S$ as then the velocity of the
magnons increases.

To conclude, we note that the quantum spin fluctuations are to a
large extent suppressed in the spin-orbital model by the
simultaneous coupling of the hole to {\it both\/} spin and orbital
excitations. In particular, they do not affect the quasiparticle
state and merely add as a small incoherent spectrum in the
incoherent high-energy part of the ladder spectrum. This is due to
the classical character of the orbitals which confine the hole
motion and prohibit its coherent motion by the coupling to the
quantum spin fluctuations. On the other hand, the hole still
couples to the spin degrees of freedom, mostly, in a classical
way, i.e., by generating string potential due to defects created
by hole motion. Thus, {\it the string} which acts on the hole
moving in the plane with AO and AF order {\it is of the composite
orbital and spin character}. This not only explains the peculiar
correspondence between the orbital and spin-orbital model but also
explains that the spins "do not hide behind the orbitals" but play
an active role in the lightly doped spin-orbital system.

\section{CONCLUSIONS}
\label{sec:conclusions}

Let us now summarize the results of the present study. We studied
the motion of single hole doped into plane of the Mott insulator
LaVO$_3$, with coexisting AF and AO order,\cite{Kha01} shown in
Fig. 1, which violates the Goodenough-Kanamori rules. The
framework to describe the hole motion in this situation is the
respective $t$-$J$ model with spin-orbital
superexchange,\cite{Kha01,Ole05} supplemented by the three-site
terms required here \cite{Dag08} (which were derived in the
Appendices) and we used this model as the starting point of our
analysis. Next, we reduced this model to the polaron-type model
and showed that the hole couples in this case simultaneously to
the collective excitations of both the AF state (magnons) and the
AO state (orbitons). The problem of finding the spectral function
was formulated and solved within the SCBA method. It should be
emphasized that, to the best of our knowledge, the problem of a
hole coupled {\it simultaneously} to the magnons and orbitons has
never been addressed before.\cite{note0}

We would like to note that the presented spectral functions, such
as the ones shown in Fig. \ref{fig:4}, predict the main features
to be measured in future photoemission experiments on the cleaved
LaVO$_3$ sample. The reader may wonder whether (apart from the
matrix-elements effects responsible for certain redistribution of
spectral intensity) any other processes, such as for example the
electron-phonon interaction, would affect hole motion to such an
extent that the spectral functions calculated here would change
qualitatively. Although we have not made any calculations for such
a more complex case so far, we suggest that they would only
enhance the ladder spectrum obtained here, since typically the
studied mechanisms are only responsible for further localization
of the hole.\cite{Kil99}

Next, it was calculated both analytically using the dominant pole
approximation and numerically using the SCBA equations that the
spectral functions contain {\it a stable quasiparticle peak, provided
the value of the superexchange $J$ was finite}. We emphasize that
this result was non-trivial as it was not clear {\it a priori\/}
whether the coupling between a hole and two excitations would lead
to a stable quasiparticle state --- e.g. the coupling between the hole
and two magnons does not lead to the stable quasiparticle peak,
cf. Ref. \onlinecite{Bal95} and Appendix \ref{app:2magnons}.
However, since the orbitons are massive excitations, the hole does
not scatter strongly on them and the quasiparticle solution
survives.

Furthermore, we studied in detail the properties of the
quasiparticle states. In particular, we looked at the differences
between the well-known spin \cite{Mar91} or orbital \cite{Bri00,
Dag08, Woh08} polarons and the obtained here spin-orbital polarons. We
checked that all of the typical quasiparticle properties of the
spin-orbital polarons such as the bandwidth, the quasiparticle
spectral weight, and the pseudogap (the distance between the
quasiparticle peak and the next excited state) are qualitatively
similar to those of the $t_{2g}$ orbital polarons studied
recently,\cite{Dag08,Woh08} and {\it it is the string picture
which dominates in the quasiparticle properties}. For example, the
bandwidth scales as $\propto J^2$ as a result of the
renormalization of the three-site terms $\propto J$, similarly as in the
purely orbital $t_{2g}$ model.\cite{Dag08,Woh08} We suggest that
the occurrence of the small but still finite bandwidth confirms
the idea of the absence of hole confinement in transition metal
oxides with orbital degeneracy presented in Ref.
\onlinecite{Dag08}.

Finally, a detailed investigation lead us to the conclusion
that the spin-orbital polarons are microscopically much more
complex than either spin or orbital polarons, and resemble the
orbital polarons only superficially. Actually, we have shown that
the spin degrees of freedom also play a significant role in the
formation of the spin-orbital polarons although their dynamics is
quenched and they are forced by the orbitals to act on the hole as
the classical Ising spins. This happens because the orbitals
confine the hole motion by forcing the hole to retrace its path
which means that the hole motion by its coupling to the quantum
spin fluctuations is prohibited. Thus, {\it the string-like
potential which acts on the hole is induced by the orbitals
although it has a joint spin-orbital character\/}.
Furthermore, it occurred that it is only in the excited part of the
spectrum that the quantum spin fluctuations contribute and are
responsible for a very small incoherent dome in the spectral function.

There is also another important feature of this orbitally induced
string formation: it could be understood as a topological effect.
This is due to the fact that it happens even if the energy of the
orbital excitations is turned to zero, i.e., when the hole moves
in the orbital sector incoherently. Hence, the mere presence of
orbitals is sufficient to obtain the (almost) classical ladder spectrum
of a hole doped into the AF and AO ordered state. This result, in
connection with the fact that the mother-compound of the
superconducting iron-pnictides shows a variety of spin-orbital
phenomena,\cite{Kru09} suggests that further investigation of
the hole propagation in spin-orbital systems is a fascinating 
subject for future studies.

Having explained all the main goals of the paper, posed in the
sixth paragraph of Sec. \ref{sec:introduction}, we can now try to
explain the paradox mentioned in the second part of the
Introduction. Actually, in the last two paragraphs above we noted
that the spin-orbital polarons resemble superficially the orbital
polarons but their nature is of joint spin and orbital character.
On the condition that this result could be also extended to the
case of the finite doping we immediately solve the paradox that
experimentally the orbitals play a significant role in the doped
cubic vanadates, although theoretically we should take into
account both spin and orbital degrees of freedom. Certainly, any
straightforward extension of the present one-hole result to the
regime of finite doping has to be rather speculative, so further
theoretical studies on the doped cubic vanadates are needed to
verify our explanation of this paradox.

\acknowledgments
We acknowledge financial support by the
Foundation for Polish Science (FNP) and by the Polish Ministry of
Science and Higher Education under Project No.~N202 068 32/1481.

\appendix
\section{Free hole motion due to the three-site terms}
\label{app:free}

\begin{figure}
\includegraphics[width=0.45\textwidth]{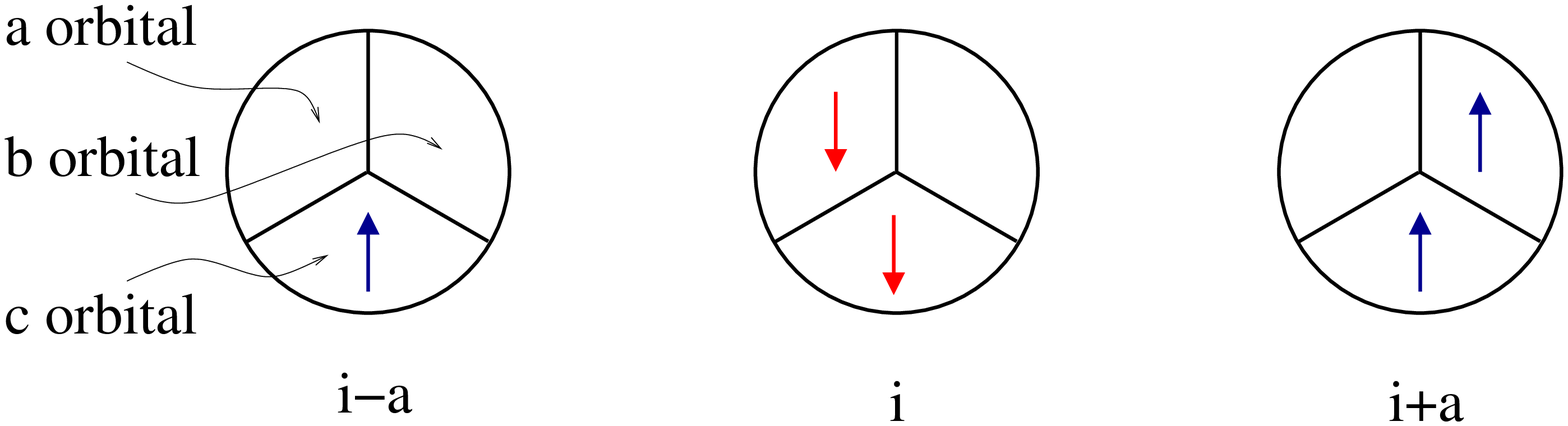}
\caption{(Color online) Three neighboring sites along the $a$
direction with such an AO and AF order that the electron on site
${\bf i}+\hat{\bf a}$ can move over the intermediate site ${\bf i}$ via the
superexchange process $\propto \tau=J/4$ to site ${\bf i}-\hat{\bf a}$
without disturbing the AO and AF order.} \label{fig:10}
\end{figure}

In what follows we derive the three-site terms which would
contribute to such a motion of a single hole that does not
disturb the AO and AF order in the ground state (as the one
described in Sec. \ref{sec:undoped}). We start our analysis by
looking at the possibility of the free hole motion along the $a$
direction, see Fig. \ref{fig:10}. We would like to move the
electron in the $b$ orbital (the one in the $a$ orbital does not
hop along this direction) from the right site to the left site
over the middle site in such a way that the spin and orbital order
present before the process stay intact. This confines the choice
of the possible high energy intermediate $d_i^3$ configurations at
the middle site to the states: (i) the ${}^4 A_2$ total spin
$|3/2,-1/2\rangle$ state with energy $U-3J_H$, and the total
$|1/2,-1/2\rangle$ spin states with (ii) ${}^2 E\frac12\theta$ and
(iii) ${}^2 E\frac12\varepsilon$ electronic configuration,
both with energy $U$. Note that all the intermediate states with
orbital singlets are excluded as they would require orbital
excitations. Thus, we arrive at the following contribution
to the free hole motion which arises due to
the three-site term processes along the $a$ direction
\begin{align}\label{eq:free1}
-\left(\frac13\frac{1}{1-3\eta} + \frac23\right) \frac{t^2}{U}\;
{\cal P}\; \tilde{b}^\dag_{{\bf i}-\hat{\bf a},\sigma} \tilde{n}_{{\bf i}a
\bar{\sigma}} \tilde{b}_{{\bf i}+\hat{\bf a},\sigma}\; {\cal P}.
\end{align}
A similar consideration but for the processes along the $b$
direction yields
\begin{align}\label{eq:free2}
-\left( \frac13 \frac{1}{1-3\eta} + \frac23 \right)
\frac{t^2}{U}\; {\cal P}\; \tilde{a}^\dag_{{\bf i}-\hat{\bf b},\sigma}
\tilde{n}_{{\bf i}b\bar{\sigma}}\tilde{a}_{{\bf i}+\hat{\bf b},\sigma}\; {\cal P}.
\end{align}
However, the processes around "the corner",\cite{Woh08} such as
for instance, first the hopping of an electron along the $b$ direction and
then along the $a$ direction in the next step, would not
contribute to the free motion. This is because an electron with a
particular ($a$ or $b$) orbital flavor can hop only along a
single cubic direction in the $ab$ plane, and thus we would have
to interchange the hopping electrons at the intermediate high
energy site, which would lead to an orbital excitation. Hence,
after adding the sums over all sites ${\bf i}$ and spins $\sigma$ and
the conjugate terms to Eqs. (\ref{eq:free1})--(\ref{eq:free2}) we
indeed end up with Eq. (\ref{eq:h3s}) presented in Sec.
\ref{sec:spin-orbital}.

Finally, one may wonder how Eqs.
(\ref{eq:free1})--(\ref{eq:free2}) could contribute to the free
hole motion since they contain four electron operators. However,
the number operators which stand in the middle of these equations
could be dropped out for the AO and AF order assumed here. More
precisely, let us e.g. assume that the intermediate site belongs
to the sublattice $A$ and concentrate on Eq. (\ref{eq:free1}).
Since we assumed in Sec. \ref{sec:undoped} that in this sublattice
the electrons have spin down and are located in the $a$ orbital
thus if, in addition, we assume that $\sigma=\uparrow$ in Eq.
(\ref{eq:free1}), then we are allowed to write $\tilde{n}_{{\bf i}a
\downarrow}\equiv 1$ and drop out this operator. Obviously, if we
assumed $\sigma=\downarrow$ or that ${\bf i}\in B$, then we would not get
any contribution. Thus, for some particular choices of $\sigma$ and
the sublattice indices Eq. (\ref{eq:free1}) would describe the
free hole motion, whereas in some other cases the hopping is
blocked and it would not contribute at all. While writing down Eq.
(\ref{eq:h3spolaron}) in Sec. \ref{sec:free} we took care of this
problem.

\section{Coupling with two magnons in the three-site terms}
\label{app:2magnons}

A careful analysis leads to the conclusion that all the three-site
terms which do not contribute to the free hole motion would lead
to the coupling between a hole and either two magnons or two
orbitons. On the one hand, since orbitons are local excitations,
see Eq. (\ref{eq:low}), the latter contribution would only
slightly enhance the string potential in the present model and
consequently the spectral functions would bear even more
signatures of the ladder spectrum. However, this effect is
expected to be quantitatively very small as the three-site terms
would contribute to the vertex as $\propto (J/4)^2$, see below,
whereas the magnon-orbiton vertices considered in the paper are of
the order of $t^2$. Thus, we can safely neglect these former
terms.

On the other hand, neglecting the terms which would lead to the
interaction between a hole and two magnons is {\it a priori\/} not
justified. We could imagine that it might lead to the hole motion
by coupling to the quantum spin fluctuations -- similarly as in
the standard spin case with the coupling between a hole and a
single magnon.\cite{Mar91} Thus, we investigated this problem in
detail by: (i) deriving the respective three-site terms, and (ii)
performing all the transformations as in Sec.\ref{sec:coupling}
which lead to the Hamiltonian written in the polaron
representation. Since all these calculations are relatively
tedious, we do not write down explicitly all the steps but merely
present the final Hamiltonian which describes the coupling between
a hole and two magnons
\begin{eqnarray}\label{eq:h2magnon}
H_{2m} &=& \frac{1}{4}\frac{\eta Jz}{1-3\eta} \; \frac{1}{2N}
\sum_{\bf k, q_1, q_2 } \left\{ V_{1y} ({\bf k, q_1, q_2})
h^\dag_{{\bf k}A}h_{{\bf \bar{k}}A}
\right. \nonumber \\
& \times &\left.(\tilde{\alpha}_{{\bf q}_{1+}}\tilde{\alpha}_{{\bf
q}_{2+}}\!+\tilde{\alpha}_{{\bf q}_{1+}}\tilde{\alpha}_{{\bf
q}_{2-}}\!-\tilde{\alpha}_{{\bf q}_{1-}}\tilde{\alpha}_{{\bf
q}_{2-}}\!-\tilde{\alpha}_{{\bf q}_{1-}}\tilde{\alpha}_{{\bf
q}_{2+}}
) \right. \nonumber \\
&+&\left. V_{1x} ({\bf k, q_1, q_2}) h^\dag_{{\bf k}B}h_{{\bf
\bar{k}}B} (\tilde{\alpha}_{{\bf q}_{1+}}\tilde{\alpha}_{{\bf
q}_{2+}}
+\tilde{\alpha}_{{\bf q}_{1+}}\tilde{\alpha}_{{\bf q}_{2-}}\right. \nonumber \\
& - &\left. \tilde{\alpha}_{{\bf q}_{1-}}\tilde{\alpha}_{{\bf
q}_{2-}} -\tilde{\alpha}_{{\bf q}_{1-}}\tilde{\alpha}_{{\bf
q}_{2+}}
)  + \mbox{H.c.} \right\} \nonumber \\
&-&\frac{1}{4}\frac{     Jz}{1-3\eta} \; \frac{1}{2N}\sum_{\bf k, q_1,
q_2 }
\left\{ V_{2y} ({\bf k, q_1, q_2}) h^\dag_{{\bf k}A}h_{{\bf \bar{k}}A}\right. \nonumber \\
&\times &\left.(\tilde{\alpha}_{{\bf q}_{1+}}\tilde{\alpha}_{{\bf
q}_{2+}}\!+\tilde{\alpha}_{{\bf q}_{1+}}\tilde{\alpha}_{{\bf
q}_{2-}}\!+\tilde{\alpha}_{{\bf q}_{1-}}\tilde{\alpha}_{{\bf
q}_{2-}}\!+\tilde{\alpha}_{{\bf q}_{1-}}\tilde{\alpha}_{{\bf
q}_{2+}}
) \right. \nonumber \\
&+&\left.V_{2x} ({\bf k, q_1, q_2}) h^\dag_{{\bf k}B}h_{{\bf
\bar{k}}B} (\tilde{\alpha}_{{\bf q}_{1+}}\tilde{\alpha}_{{\bf
q}_{2+}}
-\tilde{\alpha}_{{\bf q}_{1+}}\tilde{\alpha}_{{\bf q}_{2-}}\right.\nonumber \\
& + &\left.\tilde{\alpha}_{{\bf q}_{1-}}\tilde{\alpha}_{{\bf
q}_{2-}} -\tilde{\alpha}_{{\bf q}_{1-}}\tilde{\alpha}_{{\bf
q}_{2+}} )  + \mbox{H.c.} \right\}\,,
\end{eqnarray}
where
\begin{eqnarray}\label{eq:v1}
V_{1\mu} ({\bf k, q_1, q_2})&=&\frac12u_{{\bf q}_1}u_{{\bf q}_2}
\cos(2k_\mu-2q_{1\mu} -q_{2\mu}) \nonumber \\ &+&\frac12v_{{\bf
q}_1}v_{{\bf q}_2}  \cos(2k_\mu-q_{2\mu})\,, \\
\label{eq:v2} V_{2\mu} ({\bf k, q_1, q_2})&=&\frac14 u_{{\bf
q}_1}v_{{\bf q}_2} \cos(2k_\mu-2q_{1\mu})\,,
\end{eqnarray}
with all the symbols defined as those in Eq. (\ref{eq:htpolaron}).
Note that we neglected above all the terms of the type  $\propto
\tilde{\alpha}^\dag\tilde{\alpha}$ as they would lead to the four
boson line diagrams (the self-energies with two magnon lines are
very small, see below, and hence the self-energies with four
magnon diagrams would be even smaller).

Next, we implement the above derived processes into the SCBA
method and obtain the following equations for the additional
self-energies:
\begin{eqnarray} \label{eq:scbap1}
\Sigma^{'}_{AA}({\bf k}, \omega)\!\!&=&\!\!
\frac{z^2\lambda^2}{2N^2} \!\sum_{\bf q_1, q_2 }\!
\left\{ [\eta V_{1y}({\bf k, q_1, q_2})\!-\!V_{2y}({\bf k, q_1, q_2})]^2 \right.\nonumber \\
& \times&\!\left.\! G_{AA}({\bf k\! - q_1\! -q_2},
\omega\!+J_S\omega_{\bf q_1}\!+J_S\omega_{\bf q_2})\right.
\nonumber \\
&+& \left.\![\eta V_{1y}({\bf k, q_1, q_2})+V_{2y}({\bf k, q_1, q_2})]^2\right. \nonumber \\
& \times&\!\left.\! G_{AA}({\bf k\! - q_1\! -q_2},
\omega\!+J_S\omega_{\bf q_1}\!+J_S\omega_{\bf q_2})
\right\}, \\
\Sigma^{'}_{BB}({\bf k}, \omega)
\!\!&=&\!\!\frac{z^2\lambda^2}{2N^2} \!\sum_{\bf q_1, q_2 }\!
\left\{ [\eta V_{1x}({\bf k, q_1, q_2})\!-\!V_{2x}({\bf k, q_1, q_2})]^2\right. \nonumber \\
& \times&\!\left.\! G_{BB}({\bf k\! - q_1\! -q_2},
\omega\!+J_S\omega_{\bf q_1}\!+J_S\omega_{\bf q_2})\right.
\nonumber \\
&+& \left.\![\eta V_{1x}({\bf k, q_1, q_2})+V_{2x}({\bf k, q_1,
q_2})
]^2\right. \nonumber \\
& \times&\!\left.\! G_{BB}({\bf k\! - q_1\! -q_2},
\omega\!+J_S\omega_{\bf q_1}\!+J_S\omega_{\bf q_2}) \right\},
\label{eq:scbap2}
\end{eqnarray}
where
\begin{equation}
\lambda=\frac{1}{1-3\eta}\frac{J}{4}.
\end{equation}
This requires that we substitute for the self-energies the
following terms:
\begin{eqnarray}
\Sigma_{AA} ({\bf k}, \omega)&\rightarrow& \Sigma^{'}_{AA} ({\bf
k}, \omega)
+\Sigma_{AA} ({\bf k}, \omega), \\
\Sigma_{BB} ({\bf k}, \omega)&\rightarrow& \Sigma^{'}_{BB} ({\bf
k}, \omega) +\Sigma_{BB} ({\bf k}, \omega),
\end{eqnarray}
in the Dyson's equations (\ref{eq:dyson1})--(\ref{eq:dyson2})
which changes the SCBA equations.

Finally, we solve the modified SCBA equations on a mesh of
$16\times 16$ points. It occurs that the spectral functions
obtained with the additional self-energies Eqs.
(\ref{eq:scbap1})--(\ref{eq:scbap2}) are virtually similar to
those obtained without them (not shown). The only small difference
is a negligible enhancement of the incoherent part. This can be
understood in the following way. First, the added contributions
scale as $(J/4)^4 $ and are indeed very small as $J<t$. Second,
the vertices, given by Eqs. (\ref{eq:v1})--(\ref{eq:v2}), are
singular at e.g. ${\bf q}_1=(0,0) $ and ${\bf q}_2=(0,0)$ points.
Therefore, the self-energies associated with these vertices could
only contribute to the incoherent part of the spectrum [the
divergent vertices lead to the divergences in denominators of Eqs.
(\ref{eq:aA})--(\ref{eq:aB}) yielding the upper bounds for the
quasiparticle residues in the dominant pole approximation]. The
physical interpretation of this phenomenon is as follows: (i) the
hole produces two magnon excitation at each step it moves forward,
(ii) the hole always moves by two sites in a single step (as the
three-site terms lead to the next-nearest neighbor hopping), and
(iii) magnons "travel" in the system and cure the defects created
by the hole at a "velocity" one site per each step. Thus, one
might argue that the magnons are not "fast" enough to erase the
defects created by the hole.

In conclusion, neglecting the three-site terms which do not lead
to the free hole motion is entirely justified. Such processes are
not only quantitatively small but also they do not change the
physics qualitatively.



\begin{thebibliography}{99}
\bibitem{note1} We have verified analytically that these terms would not
                lead to any ${\bf k}$-dependence in the spectra as in the
                self-energies describing such processes one can shift the
                summation over the momenta in such a way that the
                self-energies are momentum independent (See also Sec.
                \ref{sec:analytic} for a similar calculation concerning the
                self-energies for two boson lines). Thus, these terms would
                not change the qualitative feature of the calculated spectra,
                i.e., that the ${\bf k}$-dependence of the quasiparticle
                states originates entirely from the three-site terms. Besides,
                we note that inclusion of such terms would require going
                beyond the LSW and LOW approximation, i.e., we would have to
                include the interactions between magnons and/or orbitons.
\bibitem{note2} Note, that only for didactic purpose in Secs.
                \ref{sec:spin}--\ref{sec:understanding} we calculate the
                spectral functions for $\eta=0$. However, one should bear
                in mind that such a value is unphysical and inconsistent
                with the assumed infinite value of the Hund's coupling in
                Eq. (\ref{eq:ht}).
\bibitem{note3} The obtained value of the incoherent bandwidth $W=4t$ in
                the classical limit well agrees with the retracebale path
                approximation (RPA) formula $W=4t_{\rm eff}\sqrt{l}$, where
                the effective hopping $t_{\rm eff}=t/\sqrt{2}$ due to the
                double exchange and $l=2$ is the number of possible forward
                going steps in the classical spin-orbital model, cf.
                discussion in the end of Sec. V C in Ref. \onlinecite{Woh08}.
                Note also that the narrowing of the bandwidth in the quantum
                case is due to the fact that the effective number of forward
                going steps, which the hole can make so that the spins become
                misaligned (which is the essence of RPA), is slightly reduced.
                This is because some of the spins are already overturned due
                to the quantum spin fluctuations in the full quantum model.
\end{thebibliography}
\end{document}